\documentclass[english,prd,superscriptaddress,nofootinbib,preprintnumbers,twocolumn,showpacs]{revtex4}
\usepackage[utf8]{inputenc}
\usepackage[english]{babel}
\usepackage{amsmath}
\usepackage{amsfonts}
\usepackage{amssymb}
\usepackage{epsfig}
\usepackage{graphics,psfrag,rotating}
\usepackage{graphicx}
\usepackage{dcolumn}
\usepackage{bm}
\usepackage{epstopdf}
\usepackage{color}
\usepackage[usenames,dvipsnames,svgnames]{xcolor}
\usepackage[colorlinks=true,
            linkcolor=red,
            urlcolor=gray,
            citecolor=blue]{hyperref}
  \usepackage{hyperref}
\usepackage[T1]{fontenc}
\usepackage{multirow}
\usepackage{float}

\usepackage{subfigure}

\usepackage{enumitem}

\def\3nab{\tilde{\nabla}}

\def\be {\begin{equation}}
\def\ee {\end{equation}}
\def\ba {\begin{align}}
\def\ea {\end{align}}

\def\bc {\begin{center}}
\def\ec {\end{center}}
\def\case#1/#2{\frac{#1}{#2}}

\newcommand{\bver}{\begin{verbatim}}
\newcommand{\ever}{\end{verbatim}}

\newcommand{\bea}{\begin{eqnarray}}
\newcommand{\eea}{\end{eqnarray}}
\newcommand{\beaa}{\begin{eqnarray*}}
\newcommand{\eeaa}{\end{eqnarray*}}

\newcommand{\e}{\mathrm{e}}

\def\case#1/#2{\textstyle\frac{#1}{#2}}

\def\be{\begin{equation}}
\def\ee{\end{equation}}
\def\bea{\begin{eqnarray}}
\def\eea{\end{eqnarray}}

\def\e{{\rm e}}
\def\V{\mathcal{U}}
\def\Lag{\mathcal{L}}
\def\mK{\mathcal{K}}
\def\diff{{\rm d}}
\def\gt{\tilde{g}}
\def\At{\tilde{A}}
\def\nablat{\tilde{\nabla}}

\begin{document}
%%%%%%%%%%%%%%%%%%%%%%%%%%%%%%%%%%%%%%%
\title{On the consistency of universally non-minimally coupled $f(R,T,R_{\mu\nu}T^{\mu\nu})$ theories} 
%%%%%%%%%%%%%%%%%%%%%%%%%%%%%%%%%%%%%%%
\author{Ismael Ayuso}\affiliation{Departamento de F\'{\i}sica Te\'orica I, Ciudad Universitaria, Universidad Complutense de Madrid, E-28040 Madrid, Spain.}
\author{Jose Beltr\'an Jim\'enez}
\affiliation{Centre for Cosmology, Particle Physics and Phenomenology, Institute of Mathematics and Physics, Louvain University, 2 Chemin du Cyclotron, 1348 Louvain-la-Neuve, Belgium.}
\affiliation{CPT, Aix Marseille Universit\'e, UMR 7332, 13288 Marseille,  France.}
\author{\'Alvaro\,de la Cruz-Dombriz}
\affiliation{Departamento de F\'{\i}sica Te\'orica I, Ciudad Universitaria, Universidad Complutense de Madrid, E-28040 Madrid, Spain.}


%%%%%%%%%%%%%%%%%%%%%%%%%%%%%%%%%%%%%%%
\begin{abstract}
%%%%%%%%%%%%%%%%%%%%%%%%%%%%%%%%%%%%%%%
We discuss the consistency of a recently proposed class of theories described by an arbitrary function of the Ricci scalar, the trace of the energy-momentum tensor and the contraction of the Ricci tensor with the energy-momentum tensor. We briefly discuss the limitations of including the energy-momentum tensor in the action, as it is a non fundamental quantity, but a quantity that should be derived from the action. 
 The fact that theories containing non-linear contractions of the Ricci tensor usually leads to the presence of pathologies associated with higher-order equations of motion will be shown to constrain
 the stability of this class of theories. We provide a general framework and show that the conformal  and non-minimal couplings to the matter fields usually lead to higher-order equations of motion. In order to illustrate such limitations we explicitly study the cases of a canonical scalar field, a $K$-essence field and a massive vector field. Whereas for the scalar field cases it is possible to find healthy theories, for the vector field case the presence of instabilities is unavoidable.
\end{abstract}
%%%%%%%%%%%%%%%%%%%%%%%%%%%%%%%%%%%%%%%%%
\maketitle
%%%%%%%%%%%%%%%%%%%%%%%%%%%%%%%%%%%%%%%%%
\section{Introduction} 
\label{intro} 
A key question in gravitational physics and possible extensions of Einsteinian gravity %able to address some of the open problems in Cosmology 
resides in the coupling of gravity and matter fields. Even assuming the correctness of the Equivalence Principle -- and the subsequent minimal coupling between matter and geometry  as dictated by General Relativity (GR) --  strongly supported by astrophysical and laboratory tests, violations of the minimal coupling may still be allowed in scales and at times 
where experiments have not been performed yet.
%
%For instance, high-energy theories usually predict couplings of matter to fields, such as scalar fields, and the appearance of extra degrees of freedom. %\jose{(I do not really understand this sentence here...)}
%
% Motivation
In the existing literature, there is a pleiade of proposals for non-minimal couplings 
such as those provided by scalar-tensor theories \cite{Scalar-tensor proposals}, vector-tensor theories \cite{Vector proposals} %(see also \cite{Clifton:2011jh} for a thorough review on different proposals for non-minimal couplings)
 and different couplings between matter and geometry \cite{matter-geometry couplings}, among others.
A recently proposed departing point consists of suggesting the coupling of a function of the Ricci scalar to the matter Lagrangian, a proposal which has produced numerous studies for gravitational and cosmological issues on the subject ({\it c.f.} \cite{Koivisto 2013, varia-fR-Lagrangian} and references therein).
In this work we shall consider a class of extended %modified 
gravity theories in which the gravitational action is given by a general function $f(R,T, R_{\mu\nu}T^{\mu\nu})$, where $R$ and $T$ denote the Ricci scalar and the trace of the energy-momentum tensor, respectively and $R_{\mu\nu}T^{\mu\nu}$ holds for the contraction of the Ricci and energy-momentum tensors. 
We shall herein address the theoretical consistency of this class of theories with special emphasis on couplings to either scalar or vector fields. Before proceeding, let us point out that, as it is well-known, there are important criteria 
to be fulfilled by any extended theories beyond Einsteinian gravity which would like to be claimed as a well-founded theory able to describe the gravitational interaction.
These criteria aim to guarantee the absence of instabilities such as the appearance of ghost-like modes and the exponential growth of perturbations around well-established spacetime backgrounds, among others.
%
%the violations of widely accepted energy conditions or the existence of superluminical modes,
% \jose{(I would remove the last part since the energy-conditions as origin of singularities are only well-established for GR and for non-minimally coupled scalar fields they can be violated without introducing instabilities. In the other hand, the presence of superluminal modes does not need to be a pathology. In fact,1-loop QED in curved backgrounds has superluminal modes, but it does not violate causality )}\alvaro{We can discuss on this}. 
%
For instance, an undesirable instability is the so-called Dolgov-Kawasaki instability which appears when at least one extra degree of freedom of the theory behaves as a ghost and therefore this mode would act to destabilise the theory with no stable ground state. The avoidance of the 
Dolgov-Kawasaki instability has been developed to constrain the extensively studied $f(R)$ gravity with minimal \cite{DK, Sergei_DK, DK-HuSawicki, Faraoni2007} and non-minimal couplings of the curvature with  matter \cite{Dolgov-Kawasaki-non-minimal}. 
Authors in \cite{Odintsov:2013iba} and \cite{Haghani:2013oma}
have recently addressed this instability issue for $f(R,T, R_{\mu\nu}T^{\mu\nu})$ theories. 

Another important requirement usually demanded by extended theories consists of the avoidance of the Ostrogradski instability, i.e., the fact that 
%Ostrogradski's theorem: 
a linear instability appears in Hamiltonians  associated with Lagrangians which depend upon more than one time derivative non-degenerately \cite{Ostrogradski, Woody}. Consequently this type of Hamiltonians turn out not to be bounded from below and well-defined vacuum states are absent. Theories with higher-order equations of motion can however be sensible provided that they are regarded within the framework of effective field theories. In such a framework the operators leading to the higher-order equations of motion are simply the first terms of some expansion whose adequate resummation might give rise to well-behaved theories (this is the situation for instance when one integrates heavy degrees of freedom out). Another possibility to make sense of theories with higher-order equations of motion is to remove the undesired unstable degrees of freedom from the physical spectrum of the theory or, at the classical level, constraining the physically allowed set of boundary conditions. However, one needs to make sure that such a procedure does not get spoilt by either time-evolution or coupling to other fields. This approach was followed in \cite{Jimenez:2012ak} for the case of the degenerate Pais-Uhlenbeck field.

A widely accepted way of circumventing the Ostrogradski instability when considering scalar-tensor theories of gravitation, consists of requiring the Euler-Lagrange equations to be second order even if higher order derivatives are present in the action. Following this line of reasoning, Horndeski's theorem \cite{Horndeski} provides the most general Lagrangian density for a scalar-tensor theory which provides second-order Euler-Lagrange equations, being for instance Galileons theories \cite{Galileons} subsets  of Horndeski-like theories.  Nonetheless, recent proposals have ensured second order equations 
of motion and hence the absence of Ostrogradski ghost degrees of freedom 
for theories which do not fall under the form of Horndeski-like theories. 
Among others, let us mention
 healthy theories beyond Horndeski \cite{Piazza}, the introduction of derivative couplings through a disformal metric between
the scalar and the matter degrees of freedom \cite{Zuma},  % provided that one works in the the Einstein frame.
multi-scalar  field theories \cite{MultiScalar}, 
 non-local gravity theories \cite{Non-local} 
 and non-linear combinations of purely kinetic gravity terms \cite{NonLinearHorndeski}.
% among others.

%
% f(R) theories
With sole dependency on $R$, the theories under consideration of course correspond to the extremely popular $f(R)$ gravity modified theories, which might be thought of as the only local, metric-based and generally coordinate invariant and stable modifications of gravity \cite{Woody, Non-local}. Both viability and stability conditions for $f(R)$ theories have been widely studied and guarantee the attractive character, the aforementioned avoidance of Dolgov-Kawasaki instability, the agreement with solar system tests and evolution of geodesics \cite{JCAP-Alvaro}.

%
% f(R,T) theories
With regard to Lagrangians with both $R$ and $T$ dependence, these kinds of modified gravity were originally introduced in \cite{Poplawski:2006ey} and later on considered in  \cite{fRTpaper} %by Harko {\it et al.}  
and some cosmological aspects have been already explored, such as the reconstruction of cosmological solutions \cite{Varia_fRT} 
in particular late-time acceleration ones \cite{stephaneseul3}. Also the energy conditions have been analyzed in Ref.~\cite{flavio}. The thermodynamics of Friedmann-Lema\^itre-Robertson-Walker (FLRW) spacetimes has been studied in Ref.~\cite{thermo1}. More recently the possibility of irreversible matter creation processes and the possibility of the occurrence of future singularities were addressed in \cite{Harko_August_2014} and  \cite{juliano} respectively.
Theories with non standard couplings between the geometry and the matter Lagrangian (see \cite{Nesseris:2008mq}) usually suffer from not conserving the energy-momentum tensor which implies a stringent shortcoming for their viability.  For $f(R,T)$  theories, this issue was studied in \cite{Diego-Alvaro-PRD2013}
showing that gravitational Lagrangians of the form $f_1(R)+f_2(T)$ 
can always be constructed to be consistent with the energy-momentum tensor standard conservation, at least for a single perfect fluid for adequate choices of functions $f_{1,2}$. Despite this partial success the growth rate in $f(R,T)$ theories was shown \cite{Diego-Alvaro-PRD2013} to be highly compromised by the existence of oscillations in the density contrast evolution, the occurrence of singularities and the fast growth of the density contrast in the studied models. 
%

% f(R,T,RmunuTmunu)
Theories including also terms of the form $R_{\mu\nu}T^{\mu\nu}$ have attracted some attention in recent years. Authors in  \cite{Odintsov:2013iba} claimed some motivation for these theories arguing that %due to the fact that 
the Ho\u{r}ava-like gravity power-counting renormalizable covariant gravity 
might represent the simplest power-law version of $f(R,\,T,\,R_{\mu\nu} T^{\mu\nu})$ theories \cite{Odintsov_HL}. One could then expect that such theories provide some insight between the usual approach in extended theories of gravity and Ho\u{r}ava-Lifshitz theory.
Several aspects of these theories are already available in literature. For instance,   
the energy conditions for these theories were originally  
addressed in \cite{Sharif:2013kga} where authors used models presented in \cite{Odintsov:2013iba} and \cite{Haghani:2013oma}
finding constraints on the model parameters %although the steps to derive the conditions 
from the Raychaudhuri equation.
Finally, some efforts have been made to try to establish the thermodynamics for
black holes embedded in a FLRW spacetime developing the Friedmann equations 
for spatially curved space-time and showing that for those theories  these equations can be transformed into the form of Clausius relation \cite{Thermo-Zubair}.

The paper is structured as follows: in Sec. \ref{Section:formalism} we present some generalities of the theories under consideration, paying special attention to the theoretical limitations imposed by the fact of considering the energy-momentum tensor at the level of the gravitational action. There we shall also include the multi-scalar representation which will allow us to identify the potential instabilities in a transparent way. Therein we shall specify the analysis for the restricted cases of $f(R)$ and $f(R,T)$ theories. In Sec. \ref{Section:Scalar Field} we shall focus on the case of the canonical scalar field, the appearance of instabilities for such a choice and the appropriate gravitational Lagrangians capable of preserving second order field equations either in the original formulation or in the multi-scalar representation. The same kind of studies as in the previous Section shall be performed in Secs. \ref{Section:$K$-essence Theories} and \ref{Section:Maxwell} for $K$-essence theories and vector fields respectively.  Sec. \ref{Section:Particular models} is then devoted to illustrating the shortcomings of classes of models for the gravitational Lagrangian and constraints to be imposed on the parameters in order to guarantee viability. Finally we shall end with in Sec. \ref{Section:Conclusions} with the main conclusions.
 
%%%%%%%%%%%%%%%%%%%%%%%%%%%%%%%%%%%%%%%%%%%%%%%%
%%%%%%%%%%%%%%%%%%%%%%%%%%%%%%%%%%%%%%%%%%%%%%%%%%
%
% Planck units, $(G=c=k_B =\hbar=4\pi \varepsilon_0 =1)$ will be used 
Throughout this paper, Greek indices run from 0 to 3, the symbol $\nabla$ denotes the standard covariant derivative and the signature $+,-,-,-$ is used. 
The Riemann tensor definition 
is $R^\mu_{\;\;\nu\alpha\beta}=
\partial_\alpha\Gamma^{\mu}_{\nu\beta}
-\partial_\beta\Gamma^{\mu}_{\nu\alpha}
+\Gamma^{\mu}_{\sigma\alpha}\Gamma^{\sigma}_{\nu\beta}
-\Gamma^{\mu}_{\sigma\beta}\Gamma^{\sigma}_{\nu\alpha}$.
% which coincides with the definition in has opposite sign to the one proposed in \cite{Giovannini}.}.

\section{Generalities and multi-scalar representation}
\label{Section:formalism}

The theories that we shall consider throughout this work are based on an action of the following form:
\begin{eqnarray}
S\,=\,\int{\rm d}^4 x \sqrt{-g}\Big[f(R,T,R_{\mu\nu}T^{\mu\nu})+\mathcal{L}_m(g_{\mu\nu},\Psi) \Big]\,,
\label{Action_fRT}
\end{eqnarray}
where $f$ is an arbitrary function of its arguments, $R_{\mu\nu}$ is the Ricci tensor corresponding to the Levi-Civita connection of the spacetime metric  $g_{\mu\nu}$, $R\equiv g^{\mu\nu} R_{\mu\nu}$ holds for the scalar curvature and $T^{\mu\nu}$  is the energy-momentum tensor of the matter fields $\Psi$ described by the Lagrangian $\mathcal{L}_m$ and defined as 
\begin{eqnarray}
T^{\mu\nu}\,=\,-\frac{2}{\sqrt{-g}}\frac{\delta (\sqrt{-g}\mathcal{L}_m)}{\delta g_{\mu\nu}}\,.
\label{Energy-momentum tensor}
\end{eqnarray}
Finally, $T\equiv g_{\mu\nu}T^{\mu\nu}$ holds for the trace of the energy-momentum tensor. In addition, we will further assume that the matter fields in $\mathcal{L}_m$ are minimally coupled to gravity so that all the non-minimal couplings will come from $f(R,T,R_{\mu\nu}T^{\mu\nu})$.  Since the appearance of the energy-momentum tensor in the action might be worrisome, a few words about the construction of the theories as given in (\ref{Action_fRT}) are in order here. This digression lies in the fact that the standard lore consists of defining the energy-momentum tensor as the variation of the action itself so this procedure would lead to an endless loop. One could try to construct the theory by a perturbative expansion analogous to the Gupta problem for gravity, i.e., one could start with a linear coupling of the Ricci tensor to the energy-momentum tensor of the matter fields. This will modify the energy-momentum tensor of the matter field, so the coupling will acquire a correction. This process will in general generate an infinite series that should be resummed in order to obtain the full theory. We will not pursue this approach here, but will regard (\ref{Action_fRT}) as a purely procedural way of defining the theory. In fact, theories described by (\ref{Action_fRT}) can be consistent without the infinite contributions from the non-minimal coupling to the energy-momentum tensor provided that we assume the contributions entering as arguments of the function $f$ solely correspond to the {\it energy-momentum} tensor of the matter Lagrangian alone defined through (\ref{Energy-momentum tensor}). It remains doubtful %dubious  
that one can dub energy-momentum tensor to such an object within the context of these universally non-minimally coupled theories. 
We could mention at least two reasons for this skepticism.

First of all, that object is obviously not conserved and does not correspond to the Noether current associated with infinitesimal translations. Secondly, being all fields non-minimally coupled to gravity, the difference of  $T^{\mu\nu}$ and the canonical energy-momentum tensor is not simply a total divergence\footnote{This is actually a feature present in all non-minimally coupled theories. These non-minimally couplings can be understood as arising from couplings of the graviton to surface terms \cite{Barcelo:2014mua}.}. For this particular study, we shall %simply 
consider the action (\ref{Action_fRT}) together with (\ref{Energy-momentum tensor}) as an operational approach to describe the theory. Thus we shall call energy-momentum tensor to $T^{\mu\nu}$ with the aforementioned discussed objections.

An even more worrisome aspect of theories described by (\ref{Action_fRT}) is the coupling of the Ricci tensor to the energy-momentum tensor: since $R_{\mu\nu}$ contains second derivatives of the metric tensor, and $T^{\mu\nu}$ will typically have first derivatives of the matter fields, the equations of motion are expected to be higher than second order and the Ostrogradski instability \cite{Ostrogradski} is likely to be present. It is worth stressing that it is precisely the coupling to the Ricci tensor which will generically render the theory unstable. As it is well-known, despite containing second derivatives in the Lagrangian, $f(R)$ theories avoid the Ostrogradski instability because they are constructed out of the Ricci scalar solely \cite{Woody}. However, for arbitrary functions explicitly containing the Ricci tensor, e.g. $R_{\mu\nu}R^{\mu\nu}$, a ghost associated with the higher-order derivatives arises. In fact,  in the following three Sections 
\ref{Section:Scalar Field},   \ref{Section:$K$-essence Theories} and \ref{Section:Maxwell}, 
 we shall study this feature for  cases where the matter fields are described either by scalar %(minimally and non-minimally coupled)
  or  by vector fields. We shall thus show that ghost modes are generally present in these theories due to the coupling $R_{\mu\nu}T^{\mu\nu}$ and that its avoidance considerably restricts the allowed form for the function $f$ as will then be illustrated in Section \ref{Section:Particular models}.

\subsection{Multi-scalar representation}

Before considering the presence of the Ostrogradski instabilities in this theories, we will present a general framework in which this instability (and its origin) can be more easily identified. To that end, we will rewrite the theory presented in (\ref{Action_fRT}) using the multiscalar-tensor representation. This way we intend to illustrate 
the generality with which the Ostrogradski instability will show up within these theories due to the aforementioned coupling $R_{\mu\nu}T^{\mu\nu}$. 
Lets start by rewriting action  (\ref{Action_fRT}) as
\begin{eqnarray}
S\,&=&\,\int{\rm d}^4 x \sqrt{-g}\Big[f(\chi_1,\chi_2,\chi_3)+\sum_{i=1}^3 f_{\chi_{i}}\left(P_i-\chi_{i} \right)
%
%
%R,T,R_{\mu\nu}T^{\mu\nu})
+\mathcal{L}_m 
\Big]\,\nonumber\\
&&
\label{Action_fRT_scalar-tensor representation}
\end{eqnarray}
where $\chi_{i=1,2,3}$ are auxiliary fields, $P_1 \equiv R$, $P_2 \equiv T$, $P_3 \equiv R_{\mu\nu}T^{\mu\nu}$, and $f_{\chi_j}\equiv \partial f / \partial \chi_{j}$, $j=1,2,3$. The corresponding field equations for those auxiliary fields 
are given by
\begin{eqnarray}
\frac{\partial^{2}f}{\partial\chi_i\partial\chi_j}\Big(P_i-\chi_i\Big)=0.
%&& f_{\chi_1 \chi_1}(R-\chi_1)+f_{\chi_1 \chi_2}(T-\chi_2)+f_{\chi_1 \chi_3}(R_{\mu\nu}T^{\mu\nu}-\chi_3)\,=\,0\,,\nonumber\\
%&& f_{\chi_2 \chi_1}(R-\chi_1)+f_{\chi_2 \chi_2}(T-\chi_2)+f_{\chi_2 \chi_3}(R_{\mu\nu}T^{\mu\nu}-\chi_3)\,=\,0\,,\nonumber\\
%&& f_{\chi_3 \chi_1}(R-\chi_1)+f_{\chi_3 \chi_2}(T-\chi_2)+f_{\chi_3 \chi_3}(R_{\mu\nu}T^{\mu\nu}-\chi_3)\,=\,0\,.
\label{Auxiliary fields equations}
\end{eqnarray}
Thus provided that $\det\frac{\partial^{2}f}{\partial\chi_i\partial\chi_j}\neq 0$, the only solution of
(\ref{Auxiliary fields equations})
turns out to be $\chi_1 = R$, $\chi_2=T$, $\chi_3 = R_{\mu\nu}T^{\mu\nu}$ and consequently the action (\ref{Action_fRT_scalar-tensor representation}) is dynamically equivalent to the original action (\ref{Action_fRT}). Let us stress that a necessary condition for the transformation to the multiscalar-tensor representation (\ref{Action_fRT_scalar-tensor representation}) to be valid lies in the non degeneracy, i.e., non vanishing determinant, of the matrix $\frac{\partial^{2}f}{\partial\chi_i\partial\chi_j}$. This element will play an important role later on. At this stage, we will introduce a field redefinition as follows $\varphi_i = -f_{\chi_i}$. Assuming that this redefinition is invertible, so that $\chi_i$ can be expressed  as a function of $\{\varphi_1, \varphi_2, \varphi_3\}$, the action (\ref{Action_fRT_scalar-tensor representation}) can be written as
\begin{eqnarray}
S\,&=&\,\int{\rm d}^4 x \sqrt{-g}\,\Big[\V(\varphi_1, \varphi_2, \varphi_3)-\varphi_1 R - \varphi_2 T 
\nonumber\\
&-& \varphi_3 R_{\mu\nu}T^{\mu\nu}
+\mathcal{L}_m  \Big]\,,
\label{Action_fRT_scalar-tensor representation tris}
\end{eqnarray}
where we have introduced the definition $\V(\varphi_1, \varphi_2, \varphi_3)\equiv f(\varphi_1, \varphi_2, \varphi_3)+\sum_{i=1}^{3} \varphi_i \chi_i(\varphi_1, \varphi_2, \varphi_3)$. 

Now we can follow the usual approach to disentangle the nonminimal coupling $\varphi_1 R$ by means of a conformal transformation of the form
$g_{\mu\nu} = {\rm e}^{2\Omega} \tilde{g}_{\mu\nu}$, with $\Omega= \log\frac{1}{\sqrt{16\pi G \varphi_1}}$ so the action
% $\sqrt{-g}= e^{4\Omega}\sqrt{-\tilde{g}}$
(\ref{Action_fRT_scalar-tensor representation tris}) becomes\footnote{Notice that the scalar $\Omega$ is dimensionless. To restore its natural dimension and have a canonically normalized scalar field, it should be rescaled as $\Omega\rightarrow\sqrt{\frac{8\pi G}{3}}\Omega$.}
\begin{eqnarray}
S\,&=&\,\int{\rm d}^4 x \sqrt{-\tilde{g}}\,\left\{{\rm e}^{4\Omega}\V(\Omega,\varphi_2,\varphi_3)
\right.\nonumber\\
&-&\left.\frac{1}{16\pi G}\Big(\tilde{R}-6\tilde{g}^{\alpha\beta}\partial_{\alpha}\Omega
\partial_{\beta}\Omega\Big)-\varphi_2 \tilde{T} \right.\nonumber\\
&-&\left.{\rm e}^{-2\Omega}\varphi_{3}\Big[
\tilde{R}_{\mu\nu}-2\tilde{\nabla}_{\mu}\tilde{\nabla}_{\nu}\Omega+2\tilde{\nabla}_{\mu}\Omega
\tilde{\nabla}_{\nu}\Omega \right.\nonumber\\
&-&\left. \left(2
\tilde{g}^{\alpha\beta}\partial_{\alpha}\Omega\partial_{\beta}\Omega
+\tilde{\Box}\Omega\right)\tilde{g}_{\mu\nu}
\Big]\tilde{T}^{\mu\nu}
%f(\chi_1,\chi_2,\chi_3)+f_{\chi_{1}}{\rm e}^{2\Omega}\left( \tilde{R}+Terms conf-trans -{\rm e}^{-2\Omega}\chi_1\right)+f2+f3
%
\right.\nonumber\\
&+&\left.{\rm e}^{4\Omega}\mathcal{L}_m({\rm e}^{2\Omega}\tilde{g}_{\mu\nu}, \Psi)\right\}\,,
\label{Action_fRT_scalar-tensor representation final}
\end{eqnarray}
%
%with  the well-known relation
%$\tilde R  = e^{-2\Omega}\left(R + 2(n-1)\triangle\Omega - (n-2)(n-1)\|\nabla\Omega\|^2\right) $ for n=4 becomes 
%\begin{eqnarray}
%\tilde R  = e^{-2\Omega}\left[R + 6\Box\Omega - 6(\partial\Omega)^2\right] 
%\end{eqnarray}
%
where we have have dropped a total divergence and used the transformation properties of the Ricci tensor and scalar curvature under conformal rescaling given by
\begin{eqnarray}
R_{\mu\nu}\,&=&\,\tilde{R}_{\mu\nu}-2\tilde{\nabla}_{\mu}\tilde{\nabla}_{\nu}\Omega+2\tilde{\nabla}_{\mu}\Omega
\tilde{\nabla}_{\nu}\Omega\nonumber\\
&-&\left(2
\tilde{g}^{\alpha\beta}\partial_{\alpha}\Omega\partial_{\beta}\Omega
+\tilde{\Box}\Omega\right)\tilde{g}_{\mu\nu}\,,\\
R\,&=&\,\e^{-2\Omega}\left(\tilde{R}-6\tilde{g}^{\alpha\beta}\partial_{\alpha}\Omega
\partial_{\beta}\Omega-6\tilde{\Box}\Omega\right)\,.
\end{eqnarray}
We have also defined the energy-momentum tensor in the conformally transformed frame as\footnote{Notice that this is not the energy-momentum tensor one would obtain from (\ref{Energy-momentum tensor}) in the Einstein frame, but it is defined as $\tilde{T}_{\mu\nu}\,=\,-\frac{2}{\sqrt{-\tilde{g}}}\frac{\delta (\sqrt{-g}\mathcal{L}_m)}{\delta \tilde{g}_{\mu\nu}}$. } 
\begin{eqnarray}
\tilde{T}_{\mu\nu}=\e^{2\Omega}T_{\mu\nu}\,.
\end{eqnarray}
and its trace with respect to $\gt_{\mu\nu}$ as $\tilde{T}\equiv\gt^{\mu\nu}\tilde{T}_{\mu\nu}=\e^{4\Omega}T$. After the conformal transformation we see that the degree of freedom contained in $\varphi_1$ has been transferred to the conformal mode $\Omega$. We will see later that, in some cases, the conformal transformation needs to be more general (depending also on the matter field and $\varphi_3$).

At this stage, let us further clarify the procedure sketched above when specified by several paradigmatic and simpler scenarios of extended gravity theories, such as 
Lagrangians of the forms $f(R)$ and $f(R,T)$. Immediately afterwards, we will go back to 
the general $f(R,T,R_{\mu\nu}T^{\mu\nu})$ case under study in this paper.

\vspace{0.5cm}
$\bullet$ {\it $f(R)$ case}

In the extensively studied scenario of pure $f(R)$ fourth-order gravity theories \cite{fR varia}, the previous derivation is nothing but the usual approach that clearly shows how the conformal degree of freedom behaves as a standard scalar field coupled to matter. Thus, these theories avoid the Ostrogradski instability. The non-degeneracy condition reduces in that case to $f_{RR}\neq0$. Whenever the latter condition does not hold, it simply means that the theory is linear in $R$, i.e., we are dealing with the usual Einstein-Hilbert action. 

\vspace{0.5cm}
$\bullet$ {\it $f(R,T)$ case}

For Lagrangians with an arbitrary function %$f$ 
depending only upon both $R$ and $T$, i.e., for the so-called $f(R,T)$ theories, the auxiliary field $\varphi_2$ in (\ref{Action_fRT_scalar-tensor representation final}) can be integrated out by using its own equation of motion,  which is given by
\begin{eqnarray}
\frac{\partial \V}{\partial \varphi_2}-T\,=\,0\, \Rightarrow\, \varphi_2\,=\,\varphi_2(\varphi_1,T)\,.
\label{EoM f(R,T)}
\end{eqnarray}
If the previous equation is rewritten after having performed the conformal transformation, we would obtain $\varphi_2\,=\,\varphi_2(\Omega,\tilde{T})$ instead. To proceed with our analysis, we need to  assume that this algebraic equation % \label{EoM f(R,T)}
 is in fact solvable with respect to $\varphi_2$. In some cases this does not need to be the case and, in addition, when solving equation (\ref{EoM f(R,T)}), one might find several branches corresponding to different solutions of this equation. Moreover, there is a special case when the function $\V$ is linear in $\varphi_2$ since, for such case, $\varphi_2$ acts as a Lagrange multiplier imposing a constraint equation rather than being an auxiliary field.  These cases are usually related to those models for which the auxiliary field $\chi_2$ cannot be introduced. Leaving pathological cases aside and assuming the solvability of the above equation (\ref{EoM f(R,T)}), the action (\ref{Action_fRT_scalar-tensor representation final})
for $f(R,T)$ theories after integrating out the auxiliary field $\varphi_2$ reads
\begin{eqnarray}
S\,&=&\,\int{\rm d}^4 x \sqrt{-\tilde{g}}\,\left[\e^{4\Omega}\V(\Omega,\tilde{T})+{\rm e}^{4\Omega}\mathcal{L}_m({\rm e}^{2\Omega}\tilde{g}_{\mu\nu}, \Psi)\right.\nonumber\\
&-&\left.  \frac{1}{16\pi G}\Big(\tilde{R}-6\tilde{g}^{\alpha\beta}\partial_{\alpha}\Omega
\partial_{\beta}\Omega\Big)-\varphi_2(\Omega, \tilde{T}) \tilde{T} 
\right]\nonumber\\
&=&\int{\rm d}^4 x \sqrt{-\tilde{g}}\left[ -  \frac{1}{16\pi G}\Big(\tilde{R}-6\tilde{g}^{\alpha\beta}\partial_{\alpha}\Omega
\partial_{\beta}\Omega\Big)\right.\nonumber\\
&+&\left.{\mathcal P}(\Omega,\Psi)\right],
\label{Action_f/R,T)}
\end{eqnarray}
where we have defined 
\begin{eqnarray}
{\mathcal P}(\Omega,\Psi)&\equiv&\e^{4\Omega}\V(\Omega,\tilde{T})-\varphi_2(\Omega, \tilde{T}) \tilde{T} +{\rm e}^{4\Omega}\mathcal{L}_m({\rm e}^{2\Omega}\tilde{g}_{\mu\nu}, \Psi).\nonumber\\
&&
\end{eqnarray}
 This function comprises all the matter sector terms including couplings to the conformal mode $\Omega$. Notice that the conformal mode appears with no derivatives and only the matter fields $\Psi$ will enter with derivatives in ${\mathcal P}$. In the standard case, the energy-momentum tensor will depend upon both the matter fields and their first derivatives\footnote{This is so because we are considering matter fields whose Lagrangian only depends upon first order derivatives.}. Hence the above action
with 
%From the above actionwe see that all the fields enter with first order derivatives and thus, 
these kinds of functions $f(R,T)$ will generically avoid the Ostrogradski instability except for
some pathological cases such as scenarios where for instance, expression (\ref{Auxiliary fields equations}) is not invertible or equation (\ref{EoM f(R,T)}) cannot be solved for $\varphi_2$  and therefore this construction fails. %We will illustrate this
%below for the case of a scalar field  playing the role of the matter component.
% 
The expression (\ref{Action_f/R,T)}) also tells us that
for the case of %massless 
scalar fields, the resulting Lagrangian term ${\mathcal P}$ in  
(\ref{Action_f/R,T)}) will resemble that of $K$-essence models 
and the stability conditions can then be obtained in an analogous manner to $f(R,\mathcal{L}_m)$ theories as it is done in \cite{Koivisto 2013}. It is important to keep in mind that, even though these theories can generically avoid the Ostrogradski instability, they can still have instabilities of a different nature around specific backgrounds.

% f(R,T,RT) discussion
\vspace{0.5cm}
$\bullet$ {\it General $f(R,T,R_{\mu\nu}T^{\mu\nu})$ case}

Let us now return to the general case of universally non-minimally coupled  Lagrangians of the form $f(R,T,R_{\mu\nu}T^{\mu\nu})$ as given in (\ref{Action_fRT_scalar-tensor representation final}). In this case, we can see 
the appearance of two types of problematic terms. They can be more easily identified by rewriting the action (\ref{Action_fRT_scalar-tensor representation final}) as
%\begin{widetext}
\begin{eqnarray}
S\,&=&\,\int{\rm d}^4 x \sqrt{-\tilde{g}}\,\left\{\hat{\V}(\Omega,\tilde{T},\varphi_3)\right.\nonumber\\&-&\left.  \frac{1}{16\pi G}\Big(\tilde{R}-6\tilde{g}^{\alpha\beta}\partial_{\alpha}\Omega
\partial_{\beta}\Omega\Big) \right.\nonumber\\
&-&\left.
{\rm e}^{-2\Omega}\varphi_{3}\Big[
\tilde{R}_{\mu\nu}\tilde{T}^{\mu\nu}
-\Big(2\tilde{T}^{\mu\nu}+\tilde{T}\gt^{\mu\nu}\Big)\tilde{\nabla}_{\mu}\tilde{\nabla}_{\nu}\Omega
\right.\nonumber\\
&+&\left.2\Big(\tilde{T}^{\mu\nu}-\tilde{T}\gt^{\mu\nu}\Big)\tilde{\nabla}_{\mu}\Omega\tilde{\nabla}_{\nu}\Omega
\Big]
%f(\chi_1,\chi_2,\chi_3)+f_{\chi_{1}}{\rm e}^{2\Omega}\left( \tilde{R}+Terms conf-trans -{\rm e}^{-2\Omega}\chi_1\right)+f2+f3
%
+{\rm e}^{4\Omega}\mathcal{L}_m({\rm e}^{2\Omega}\tilde{g}_{\mu\nu}, \Psi)\right\} \,,\nonumber\\
\label{Action_fRT_scalar-tensor alternative}
\end{eqnarray}
%\end{widetext}
where, in the very same manner as for the $f(R,T)$ case, we have integrated out
the auxiliary field $\varphi_2$ by using its own equation of motion and we have rearranged terms including $\varphi_2$ in %\alvaro{analogous. What does it mean?} 
$\hat{\V}(\Omega,\tilde{T},\varphi_3)\equiv\e^{4\Omega}\V-\varphi_2(\Omega,\tilde{T},\varphi_3)\tilde{T}$. With the action expressed in this form we can clearly identify therein
two potential stability problems 
%the aforementioned problems 
that we describe in the following:

\begin{itemize}
\item[1.-] On the one hand, there are terms with second order derivatives of
 the conformal mode of the form $\mK^{\mu\nu}\tilde{\nabla}_{\mu}\tilde{\nabla}_{\nu}\Omega$ where $\mK^{\mu\nu}$ contains first derivatives of the matter fields. Therefore, this term will lead to higher-order equations of motion and, thus, the propagation of additional degrees of freedom which will correspond to unstable Ostrogadski modes. As we will discuss below in more detail, there are cases in which the structure of $\mK^{\mu\nu}$ makes it possible to avoid higher-order equations of motion. This opens the possibility of having consistent theories of the discussed type, but
the universal validity of theories with non-minimal couplings must be abandoned, as we will see later. For instance 
we shall show how standard vector field theories do lead to Ostrogradski modes.

\item[2.-] On the other hand, there is also 
a non-minimal coupling of the Ricci tensor to the energy-momentum tensor which might 
be the origin of additional instability problems. In particular:

\begin{itemize}
\item[2.1.-] \label{2.1} For a fixed curved background, this coupling will modify the kinetic term of the matter field
and could turn it into a ghost (and/or other type of instabilities)  due to the non-definite signature of the Ricci tensor. 

\item[2.2.-] For dynamical gravitational fields, these non-minimal couplings will generally introduce additional propagating degrees of freedom associated with higher-order 
equations of motion with the corresponding Ostrogradski instability. Again in this case, for specific types of matter and particular choices of the function $f$, these non-minimal couplings might actually be stable.

\end{itemize}
\end{itemize}
The two points above are general features of these theories, but there can be ways of avoiding %getting around 
the Ostrogradski instabilities. In the general case, we also have the auxiliary field $\varphi_3$  that can be integrated out after having used its equation of motion, given by
\begin{eqnarray}
\label{EoM f(R,T,RT) varphi3}
&&
{\rm e}^{4\Omega}\frac{\partial \V(\Omega,\varphi_2,\varphi_3)}{\partial \varphi_3}
-\e^{-2\Omega}
\Big[
\tilde{R}_{\mu\nu}-2\tilde{\nabla}_{\mu}\tilde{\nabla}_{\nu}\Omega\nonumber\\&&+\,2\tilde{\nabla}_{\mu}\Omega
\tilde{\nabla}_{\nu}\Omega
-\left(2
\tilde{g}^{\alpha\beta}\partial_{\alpha}\Omega\partial_{\beta}\Omega
+\tilde{\Box}\Omega\right)\tilde{g}_{\mu\nu}
\Big]\tilde{T}^{\mu\nu}\,=\,0.\nonumber\\
\label{multiscalarAlt}
\end{eqnarray}
Here one should bear in mind the same subtleties as discussed after Eq. (\ref{EoM f(R,T)}) for the case of $\varphi_2$. Assuming again that the above equation can be algebraically solved for $\varphi_3$, the solution can be plugged back into the action (\ref{Action_fRT_scalar-tensor alternative}) to remove the dependence on $\varphi_3$. To be more precise, when both $\varphi_2$ and $\varphi_3$ are present, the condition for the solvability of $\varphi_2$ and $\varphi_3$ in terms of their own equations of motion is given by the non-degeneracy of the matrix $\partial^{2}\mathcal{U}/\partial\varphi_i\partial \varphi_j$ with $i,j=2,3$. In other words, this will be the condition for such fields to be actual auxiliary fields and not lagrange multipliers. This condition is actually linked to the condition for the validity of the Legendre transformation of the original action because one can easily show that
\be
\frac{\partial^{2}\mathcal{U}}{\partial\varphi_i\partial\varphi_j}=-\left(\frac{\partial^{2}f}{\partial\chi_i\partial\chi_j}\right)^{-1}
\ee
again, for $i,j=2,3$. The fact that $\varphi_3$ is an auxiliary field will be a crucial obstruction for the stability of the theory 
as we shall exemplify below in the subsequent Sections. In fact, this shortcoming will motivate the use of gravitational Lagrangians which are linear in $R_{\mu\nu}T^{\mu\nu}$ so that 
the system (\ref{Auxiliary fields equations}) is degenerate and $\varphi_3$ cannot be defined as an auxiliary field. This will make contact with Horndeski-type of interactions because provided that $\varphi_3$ is not really an auxiliary field that needs to be integrated out, we see from (\ref{multiscalarAlt}) that the potentially dangerous term for the conformal factor is linear in the second derivatives of the conformal mode $\Omega$, i.e., it is of the form ${\mathcal K^{\mu\nu}}\tilde{\nabla}_{\mu}\tilde{\nabla}_{\nu}\Omega$ where ${\mathcal K}^{\mu\nu}$ only depends on derivatives of the matter fields. Thus, it is a general result that provided ${\mathcal K}^{\mu\nu}$ does not contain time derivatives\footnote{Actually, if $\mathcal{K}^{\mu\nu}$ only depends on $\Omega$, even if it contains time derivatives, the equations of motion are second order. See Appendix for more details.}, the field equations are actually second order. Consequently, the Ostrogradski instability is avoided. For this purpose, it is crucial that $\varphi_3$ is not an auxiliary field since otherwise, even if the structure of ${\mathcal K^{\mu\nu}}\tilde{\nabla}_{\mu}\tilde{\nabla}_{\nu}\Omega$ is correct, after integrating out $\varphi_3$, the avoidance of the Ostrogradski instability will be compromised.

Upcoming sections shall in fact be devoted to better explaining % 
and illustrating the aforementioned general statements when applied to specific choices of the matter Lagrangian, in particular to scalar and vector fields.

% (\ref{Action_fRT_scalar-tensor representation final})

%   {\bf Ideas after Jose's discussion}
%   \begin{enumerate}
%  \item[1.] If $\det \neq 0$ then the theory possesses Ostrodgradski instability 
%  \item[2.]  If the theory is Horndeski-like then the $det =0$ and the method is not valid.
%  \item[3.] $1.$ and $2.$ combined tell that  If $\det \neq 0$ then the theory is not Horndeski-like and Ostrogradski instability is present.
%  \end{enumerate}
%

%%%%%%%%%%%%%%%%%%%%%%%%%%%%%%%%%%%%%%%%%%%%%%%%%%%%%

\section{Canonical scalar Field }
\label{Section:Scalar Field}
In order to clarify the general discussion performed in the previous section we shall focus here on a set-up where the matter sector is given by
a scalar field.
Thence we start by considering the simplest case of a canonical scalar field and we will  make contact with Horndeski type of interactions to guarantee 
the avoidance of Ostrogradski instability. Thus, the matter Lagrangian will be given by
\begin{eqnarray}
\mathcal{L}_m\,=\,\Lag_\phi=\frac{1}{2}\partial_\mu\phi \partial^\mu\phi-V(\phi)\,,
\label{Scalar field Lagrangian}
\end{eqnarray}
with $V(\phi)$ the potential of the scalar field. The corresponding energy-momentum tensor for the scalar field is
\be
T_{\mu\nu}\,=\,\partial_{\mu}\phi\,\partial_{\nu}\phi-g_{\mu\nu}\mathcal{L}_\phi\,.
\label{Tmunu Scalar field}
\ee
and the trace of the energy-momentum tensor and the contraction $R_{\mu\nu}T^{\mu\nu}$ provide
\begin{eqnarray}
T\,&=&\,-(\partial\phi)^2+4V(\phi)\,,\nonumber\\
R_{\mu\nu}T^{\mu\nu}\,&=&\,G_{\mu\nu}\partial^{\mu}\phi\partial^{\nu}\phi+R\, V(\phi)\,,
\label{Contractions Scalar field}
\end{eqnarray}
with the usual definition for the Einstein tensor $G_{\mu\nu}\equiv R_{\mu\nu}-1/2 g_{\mu\nu}R$ and we have used the notation $(\partial \phi)^2\equiv\partial_{\mu}\phi\,\partial^{\mu}\phi$. Once the arguments of the gravitational Lagrangian have been expressed in terms of the scalar field, action (\ref{Action_fRT}) will take the form
\begin{eqnarray}
S\,&=&\,\int {\rm d}^4 x \sqrt{-g} \Big[f\Big(R, -(\partial\phi)^2+4V(\phi), G^{\mu\nu} \partial_{\mu}\phi \partial_{\nu}\phi\nonumber\\
&+&R V(\phi)
\Big) + \mathcal{L}_{\phi}(g_{\mu\nu}, \phi)\Big]\,,
\label{Action Canonical Scalar Field}
\end{eqnarray}
so that we obtain a non-minimally coupled theory with derivative couplings of the scalar field to the Ricci curvature. As it is well-known, this type of couplings leads to higher-order equations of motion and consequently they generally 
suffer from the Ostrogradski instability \cite{Ostrogradski}. This instability is actually present in general theories containing arbitrary contractions of the Riemann tensor, being the $f(R)$ theories an exceptional case
where the presence of constraints %(or )
removes the unstable dynamical degree of freedom\footnote{This fact is a consequence of the degeneracy of the transformation to canonical variables, which is a crucial step in the Ostrogradski construction \cite{Ostrogradski}.} as explained in the previous Section.

The usual approach to avoid this instability a priori %from the very beginning 
consists of building actions  leading to second-order equations of motion. For instance, in a purely gravitational context
such actions are given by the Lovelock invariants which in four dimensions reduce to a cosmological constant, the Ricci scalar and the Gauss-Bonnet term, the latter being a topological invariant.
In this realm, in the context of scalar-tensor theories, the analogous Lagrangians 
with second order equations of motion were obtained by Horndeski \cite{Horndeski}. The appropriate gravitational Lagrangian can be written as a sum of the following four terms:
\begin{eqnarray}
\mathcal{L}_2&=&K(\phi,X)\,,\label{L2}\\
\mathcal{L}_3&=&G_3(\phi,X)\Box\phi\,,\label{L3}\\
\mathcal{L}_4&=&G_4(\phi,X)R\nonumber\\
&&-\,G_{4,X}(\phi,X)\Big[(\Box\phi)^2-(\nabla_\mu\nabla_\nu\phi)^2\Big],%\nonumber\\
%&& \,
\label{L4}\\
\mathcal{L}_5&=&G_5(\phi,X)G_{\mu\nu}\nabla^\mu\nabla^\nu\phi+\frac16G_{5,X}(\phi,X)\Big[(\Box\phi)^3\nonumber\\
&&-\,3(\Box\phi)(\nabla_\mu\nabla_\nu\phi)^2+2(\nabla_\mu\nabla_\nu\phi)^3\Big]\,,\label{L5}
%\label{Horndeski-Jose}
\end{eqnarray}
where $X\equiv \frac{1}{2}\partial_{\mu}\phi\,\partial^{\mu}\phi$, $K$ and $G_{3,4,5}$ are arbitrary functions of $\phi$ and $X$ and the subindex $,X$ refers to the derivative with respect to $X$.
Therefore, in order to avoid Ostrogradski instability for the theories considered in (\ref{Action_fRT}), it is sufficient to guarantee that the action lies within the aforementioned Horndeski theories either for the original form of the action or in the multiscalar representation. Notice that in the multiscalar representation we will eventually have two scalar fields (the matter scalar field plus the conformal mode), so the considered actions will actually be more general than the above Horndeski terms. We clearly see 
that this requirement will extremely constrain the permitted form of the gravitational Lagrangians $f(R,T,R_{\mu\nu}T^{\mu\nu})$. We should stress here that, while the Horndeski terms are the most general ones explicitly leading to second order equations of motion, they are not the most general theories that propagate one spin-2 plus one spin-1 fields. A more general class of theories has been shown to propagate exactly the same degrees of freedom as the Horndeski terms even though the equations are of a higher order  \cite{Piazza}. The reason can be traced to the existence of {\it hidden} constraints that reduce the required number of boundary conditions. Some terms within that class of theories can actually be related to some Horndeski terms by means of a general disformal transformation  \cite{Zuma}.

\subsection{Preservation of second order field equations}
\label{subsubsection:Preservation}
We will first look for conditions on the function $f$ so that the theory contains no higher than second order equations of motion in its original form. This will guarantee that neither the gravitational sector nor the matter sector will propagate more degrees of freedom than it corresponds, i.e., the two corresponding to the graviton plus one associated with the scalar field (or, in other words, the conformal mode is not excited). In the next subsection we will drop this condition to let the conformal mode propagate as well and we will find conditions for the absence of Ostrogradski instabilities also in that case. 

From  the form of the Horndeski Lagrangians, one can a priori infer that curvature-scalar field couplings need to be linear in the curvature according to $\mathcal{L}_{4,5}$ to maintain the second order nature of the field equations. This linearity in the curvature implies a first stringent constraint on the function $f$, which consequently needs to be of the form 
\begin{eqnarray}
f(R,T,R_{\mu\nu}T^{\mu\nu})=f_{1}(T)R+f_{2}(T)R_{\mu\nu}T^{\mu\nu} + f_3(T)\,,\nonumber\\
\label{First constraint}
\end{eqnarray}
for arbitrary functions $f_{1,2,3}(T)$. Of course this is not the most general case free from the Ostrogradski instability because it is well-known that higher-order terms in derivatives can lead to stable field equations provided they correspond to the special class of degenerate theories. For instance, we could have added an arbitrary function of the Ricci scalar in (\ref{First constraint}) or considered more general scalar-tensor interactions as 
commented above,  without introducing this instability. Let us also remember that $f_3(T)$ is in fact a function of $X$ and $\phi$ and therefore lies in $\mathcal{L}_2$.

For our scalar field scenario with the result in 
(\ref{Contractions Scalar field}) and the form for the function $f$ given in (\ref{First constraint}), the corresponding action becomes
\begin{eqnarray}
S&=&\int\diff^4x\sqrt{-g}\Big[\left(f_1(T)+f_2(T) V\right)R\nonumber\\
&+&f_{2}(T)G^{\mu\nu}\partial_{\mu}\phi\,\partial_{\nu}\phi+f_3(T)+\Lag_\phi\Big]\,,
\label{function Scalar Field Horndeski}
\end{eqnarray}
with $f_{1,2,3}$ functions only of $T= -2 X + 4 V(\phi)$. Now we want to obtain further constraints on $f_{1,2}$ so that the action can be mapped into
Horndeski-like terms. The absence of terms of the type $(\nabla\nabla\phi)^3$ in (\ref{function Scalar Field Horndeski}) suggests that $f_2$ should be 
a function of $\phi$ only and not of $X$. 
%Thus the functions $f_{1,2}$ must be mapped to the Horndeski terms (\ref{L2})-(\ref{L5}).
%$G(\phi,\,X)$ 
%In order to proceed, let us rewrite the second term in the r.h.s. of (\ref{function Scalar Field Horndeski}) as follows,
%
%\begin{eqnarray}
%f_{2}G^{\mu\nu}\partial_{\mu}\phi\,\partial_{\nu}\phi\,=\, f_2\left[
%(\Box\phi)^2-(\nabla_\mu\nabla_\nu\phi)^2\right]-\frac{1}{2}f_2\left(\partial\phi\right)^{2}R+\left[(\partial_{\alpha}f_2)(\partial^{\alpha}\phi) g^{\mu\nu}-(\partial^{\mu}f_2)(\partial^{\nu}\phi)\right]\nabla_{\mu}\nabla_{\nu}\phi \,.
%\label{Simplification_1}
%\end{eqnarray}
%
%where the last term in the previous expression will typically lead to higher-order (beyond second) derivatives unless $f_2$ is just a function of the field $\phi$. 
However $f_2$ is a function of the energy-momentum tensor trace $T=-2X+4V(\phi)$ so that it also explicitly depends upon $X$. Consequently the only possibility left is that $f_2$ is simply a constant. Thus the only remaining arbitrary function would be $f_1(T)$. Nonetheless, this function is not arbitrary either since if (\ref{function Scalar Field Horndeski}) is required to be Horndeski-like, the only way the first term in (\ref{function Scalar Field Horndeski}) can be mapped to $G_4$ is for $f_1+f_2V$ being solely a function of $\phi$, which analogously to the reasoning used for $f_2$, leads to conclude that $f_1$ also needs to be a constant. Therefore, the requirement explicitly guaranteeing second order field equations in $f(R,T,R_{\mu\nu} T^{\mu\nu})$ theories with a standard scalar field as matter sector leads to actions of the form
\begin{eqnarray}
S&=&\int\diff^4x\sqrt{-g}\left[\Big(c_1+c_2V(\phi)\Big)R\right.\nonumber\\
&+&\left.\frac12\Big(g^{\mu\nu}+c_2G^{\mu\nu}\Big)\partial_{\mu}\phi\,\partial_{\nu}\phi-V(\phi)+f_3(-2X+4V)\right]\nonumber\\
\label{Action_Scalar_Field}
\end{eqnarray}
with $c_{1,2}$ some constants\footnote{The curly brackets after $f_3$ must be understood as the argument of $f_3$ since $T=-2X+4V$ in this case according to (\ref{Contractions Scalar field}).}. One can now immediately identify the different terms in the above action with the corresponding Horndeski Lagrangians with
\begin{eqnarray}
&&G_2=\frac{1}{2}g^{\mu\nu}\partial_\mu\phi\partial_\nu\phi-V(\phi)+f_3\left(2X+4V(\phi)\right),\nonumber\\
&&G_3=0,\nonumber\\
&&G_4=c_1+c_2V(\phi),\nonumber\\
&&G_5=-c_2\phi,
\end{eqnarray}
where we have used that, via integration by parts, $G^{\alpha\beta}\partial_\alpha\phi\partial_\beta\phi\rightarrow-\phi  G^{\alpha\beta}\nabla_\alpha\nabla_\beta\phi$.
For a scalar field without potential, i.e., with a shift symmetry $\phi\rightarrow\phi+c$ with $c$ a constant, the first term in (\ref{Action_Scalar_Field}) simply renders the Einstein-Hilbert term (and we should identify $c_1\equiv-(16\pi G)^{-1}$) whereas the last term gives a contribution in the form of a $K$-essence term. 
If we further set $f_3=0$, then we end up with a non-minimally derivatively coupled scalar field whose non-minimal coupling is to the Einstein-tensor. This simplified case was explored in \cite{Germani:2010hd} as a model of inflation and  in \cite{Rinaldi:2012vy} black-hole solutions were obtained. The non-minimal derivative coupling to the Einstein tensor also arises in the covariantization of the decoupling limit of massive gravity  \cite{deRham:2011by}. Thus, these models can be regarded as specific cases of the general $f(R,T,R_{\mu\nu}T^{\mu\nu})$ theories where the function is subject to be simply $f=R_{\mu\nu}T^{\mu\nu}$ or, in other words, the derivative coupling to the Einstein tensor can be alternatively seen as a coupling of the Ricci tensor to the energy-momentum tensor of the scalar field.

%\vspace{1cm}
%{\bf Prospects??}

\subsection{Multiscalar-tensor representation analysis}
\label{sub subsection:Multiscalar-tensor representation analysis}

Let us now consider more general actions by using the multi-scalar representation for the canonical scalar field studied above, i.e., we will now let the conformal mode be excited. This approach will enable us to detect and identify the instabilities that 
$f(R,T,R_{\mu\nu}T^{\mu\nu})$ Lagrangians may suffer for more general models with field equations beyond second order. Equivalently, this procedure will allow us to find general actions with higher-order field equations which are actually {\it healthy} in a similar manner to $f(R)$ theories. Before proceeding, a subtlety should be remarked on at this stage: as we can see from (\ref{Contractions Scalar field}), the coupling $R_{\mu\nu}T^{\mu\nu}$ will also generate a coupling between the scalar field potential and the Ricci scalar $V(\phi)R$. Therefore, after performing the conformal transformation, the definition of the conformal mode needs to be modified to $\Omega\equiv\log\frac{1}{\sqrt{16\pi G(\varphi_1+V(\phi)\varphi_3)}}$. After taking this into account, the action becomes\footnote{In the expression (\ref{action_Scalar_Field_multiscalar}) it must be understood that indices are now raised and lowered with the metric $\tilde{g}_{\mu\nu}$. For instance, $(\partial\Omega)^2\equiv\gt^{\alpha\beta}\partial_\alpha\Omega\partial_\beta\Omega$.}
\begin{eqnarray}
S&=&\int {\rm d}^4x\sqrt{-\tilde{g}}\left\{
\hat{\V}(\Omega,\tilde{T},\varphi_3)-\frac{1}{16\pi G}\left[\tilde{R}-6(\partial\Omega)^2\right]
\right.\nonumber\\&&\left.
-\,\varphi_3\left[\tilde{G}^{\mu\nu}\partial_{\mu}\phi \partial_{\nu}\phi
+2\left(\partial_{\mu}\Omega\partial^{\mu}\phi\right)^2
+(\partial\Omega)^2(\partial\phi)^2\right.\right.\nonumber\\
&+&\left.\left.2\Big(\tilde{g}^{\mu\nu}(\partial\phi)^2-\partial^{\mu}\phi\partial^{\nu}\phi%-3\e^{2\Omega}V\gt^{\mu\nu}
\Big)\tilde{\nabla}_{\mu}\tilde{\nabla}_{\nu}\Omega
\right]
\right\}\,.
\label{action_Scalar_Field_multiscalar}
\end{eqnarray}
In this case, the last expression explicitly shows why the fact that $\varphi_3$ is an auxiliary field will be problematic. Although the terms in brackets in the second line of the above expression provide the {\it appropriate} structure in order to guarantee second-order equations of motion, one will in general generate potentially 
dangerous terms again after integrating $\varphi_3$ out. 
For instance,  for a theory with second derivatives of a scalar field (the conformal mode in our case) one typically needs the second derivatives to appear linearly, as it happens in (\ref{action_Scalar_Field_multiscalar}). However, after integrating $\varphi_3$ out, such second derivatives will enter non-linearly in the action, thus spoiling the required structure. A loophole in this conclusion occurs for 
a Lagrangian $f$ with a linear dependence on the argument $R_{\mu\nu}T^{\mu\nu}$. In such a case, the auxiliary field $\varphi_3$ cannot be defined and the resulting action would read
\begin{eqnarray}
S&=&\int {\rm d}^4x\sqrt{-\tilde{g}}\left\{ 
\hat{\V}(\Omega,\tilde{T},\varphi_3)-\frac{1}{16\pi G}\left[\tilde{R}-6(\partial\Omega)^2\right]
\right.\nonumber\\
&-&\left.
\alpha\left[\tilde{G}^{\mu\nu}\partial_{\mu}\phi \partial_{\nu}\phi
+2\left(\partial_{\mu}\Omega\partial^{\mu}\phi\right)^2
+(\partial\Omega)^2(\partial\phi)^2 \right.\right.\nonumber\\
&+&\left.\left. 2\Big(\tilde{g}^{\mu\nu}(\partial\phi)^2-\partial^{\mu}\phi\partial^{\nu}\phi%-3\e^{2\Omega}V\gt^{\mu\nu}
\Big)\tilde{\nabla}_{\mu}\tilde{\nabla}_{\nu}\Omega
\right]
\right\}\,
\label{action_Scalar_Field_multiscalar2}
\end{eqnarray}
with $\alpha$ a constant parameter. The potentially dangerous terms are in the second line of the above expression. The non-minimal coupling of the field $\phi$ is to the Einstein tensor and multiplied by a function of only $\Omega$ and not its derivatives. Therefore, this coupling will be of the Horndeski form. The other term that can potentially lead to higher-order equations of motion is the coupling of $\phi$ to the second derivatives of $\Omega$. However, this is also a safe interaction because the tensor structure of $\mathcal{K}^{\mu\nu}\equiv\tilde{g}^{\mu\nu}(\partial\phi)^2-\partial^{\mu}\phi\partial^{\nu}\phi$ is such that $\mathcal{K}^{00}$ does not contain any time derivatives and this guarantees that the field equations will be second order (see Appendix for further details).

\section{$K$-essence Theories }
\label{Section:$K$-essence Theories}
After studying in detail the case of a canonical scalar field, let us move on to a more general case where the action for the scalar field is given by a $K$-essence model, i.e., the matter Lagrangian now reads
\begin{eqnarray}
\mathcal{L}_m=\mathcal{L}_{K}=K(\phi, X),
\end{eqnarray}
where $K(\phi, X)$ is an arbitrary function of its arguments. The previous case for a canonical scalar field corresponds to the particular case $K(\phi, X)=X-V(\phi)$. We will proceed in a similar manner to the precedent Section and, in addition to obtain constraints on the gravitational action $f$, we will obtain conditions on $K$. However, here we will aim to show that the freedom in the choice for the function $K(\phi,X)$ allows us for more general functions $f(R,T,R_{\mu\nu}T^{\mu\nu})$. The relevant quantities in this case are given by
\begin{eqnarray}
&& T_{\mu\nu}=K_{,X}\partial_{\mu}\phi\partial_{\nu}\phi-g_{\mu\nu}K,\label{Kessence1} \\
&& T=2K_{,X}X-4K,\label{Kessence2} \\
&& R_{\mu\nu}T^{\mu\nu}=K_{,X}G^{\alpha\beta}\partial_\alpha\phi\partial_\beta\phi+R\left(K_{,X}X-K\right),
\label{Kessence3}
\end{eqnarray}
where we can start to see the role that the form of the $K$-function and its dependence on $X$ might play in the coupling terms. In particular, we see that the coupling $R_{\mu\nu} T^{\mu\nu}$ generates a derivative interaction with the Ricci scalar in addition to the non-minimal coupling to the Einstein tensor that we obtained in the canonical scalar field case. In fact, this might be the origin of pathologies for general $K$-essence models.
The general action will take the form:
\begin{eqnarray}
S&=&\int{\rm d}^4 x \sqrt{-g}\,\left\{f\left(R,\,2 K_{,X}g^{\mu\nu}\partial_{\mu}\phi\partial_{\nu}\phi-4K(X,\phi),\right.\right.\nonumber\\
&&\left.\left.\,2 K_{,X} G^{\alpha\beta}\partial_\alpha\phi\partial_\beta\phi+R\left[ K_{,X}X-K(\phi, X)\right]\right)\right.\nonumber\\
&+&\left.K(\phi,X)\right\}.
\label{Action $K$-essence}
\end{eqnarray}

Analogously to the scheme in the previous section, we will look for conditions to be imposed on the function $f$ so that the theory does not lead to higher than second order equations of motion in its original form. Again, for the action to be of the Horndeski type, the non-minimal coupling must be linear in the curvature and, therefore, the function $f$ should also be linear in $R$ and $R_{\mu\nu}T^{\mu\nu}$. After imposing such restrictions, the action 
(\ref{Action $K$-essence}) simply reads
\begin{eqnarray}
S&=&\int{\rm d}^4 x \sqrt{-g}\,\Big[f_1(T)R+f_2(T)+f_3(T)R_{\mu\nu}T^{\mu\nu}\nonumber\\&+&K(\phi,X)\Big]\nonumber\\
&=&\int{\rm d}^4 x \sqrt{-g}\,\Big\{\left[f_1(T)+f_3(T)\Big(K_{,X}X-K\Big)\right]R\nonumber\\&+&f_2(T)+f_3(T)K_{,X}G^{\mu\nu}\partial_\mu\phi\partial_\nu\phi+K(\phi,X)\Big\}\,,
\label{Action $K$ essence bis}
\end{eqnarray}
with $T$ given by (\ref{Kessence2}).  A reasoning similar to the one below (\ref{function Scalar Field Horndeski}) allows us to establish that $f_3K_{,X}$ should be a function of $\phi$ only and not of $X$, i.e., 
\begin{eqnarray}
\partial_X\Big(f_3K_{,X}\Big)=0\,,
\label{Kessencecondition1}
\end{eqnarray}
 so that $f_3K_{,X}= g_1(\phi)$. Again, proceeding analogously to the reasoning below (\ref{function Scalar Field Horndeski}), we can also conclude that $f_1(T)+f_3(T)\Big(K_{,X}X-K\Big)$ should also be a function of only $\phi$. Now, if we combine these two conditions we finally obtain that the following equation must be fulfilled 
\be
\partial_X\Big[f_1(T)+g_1(\phi)X-f_3K\Big]=0\,,
\ee
which, by using that $\partial_Xf_i=2f'_i\Big(K_{,XX}X-K_{,X}\Big)$ for $i=1,3$, can be expressed as
\begin{eqnarray}
2\Big(f_1'&-&f_3'K\Big)XK_{,XX}-2\Big(f_1'+f_3\Big)K_{,X}+f_3'\big(K^2\big)_{,X}\nonumber\\
&=&-g_1,
\label{Kessencecondition}
\end{eqnarray}
where primes denote derivative with respect to the argument ($T$ in this case). We see that for the canonical scalar field, for which $K=X-V(\phi)$,  the above conditions imply that $f_1$ and $f_3$ can only depend on $\phi$ so that, being functions of $T$, this is only possible if they are indeed constant functions, in agreement with our previous result. In the present case however, the freedom in the choice of the function $K(\phi,X)$ allows for more general cases provided Eq. 
(\ref{Kessencecondition1}) together with (\ref{Kessencecondition}). For instance, a straightforward generalization of the canonical scalar field is to impose $K_{,XX}=0$ or, more explicitly, a model with
\begin{eqnarray}
K(\phi,X)=h(\phi)X-V(\phi)\,,
\label{$K$-essence form Horn}
\end{eqnarray}
where $h(\phi)$ and $V(\phi)$ are arbitrary functions of the scalar field. According to (\ref{Kessencecondition1}), this condition also implies that $f_3$ must be a constant  and, additionally, Eq. (\ref{Kessencecondition}) further imposes that $f_1$ also needs to be a constant. Notice that for this particular case, we have that $K_{,X}X-K=V(\phi)$ and, therefore, all the non-minimal derivative couplings in the action are through the functions $f_1(T)$ and $f_3(T)$, as in the canonical scalar field case.
%
%Accordingly,  the identification with Horndeski-like term $G_4$ is straightforward being $G_4=f_1(T)+f_3(T)\left(K_{,X}X-K\right)$ which does depend on $X$. Therefore we need to the second term appearing on $\mathcal{L}_4$ in (\ref{L4}). However this term is absent in (\ref{Action $K$ essence bis}) which forces us to conclude that the only allowed term multiplying $R$ in the previous action needs to be a constant, i. e., $f_1(T)=\alpha$, $f_3(T)=\beta$ and $K_{,X}X-K=g(\phi)$. The latter condition is actually a constraint on the function $K$ that needs to be of the form:
 Thus, our action for (\ref{$K$-essence form Horn}) with $f_1$ and $f_3$ constants will adopt the following form:
\begin{eqnarray}
S&=&\int{\rm d}^4 x \sqrt{-g}\,\left\{\hat{g}(\phi)R+\hat{h}(\phi)G^{\mu\nu}\partial_\mu\phi\partial_\nu\phi+\hat{K}(\phi,X)\right\}\,,\nonumber\\
&&
\end{eqnarray}
where $\hat{g}(\phi)=f_1+f_3g(\phi)$, $\hat{h}(\phi)=f_3 h(\phi)$ and $\hat{K}(\phi,X)\equiv f_2(T)+K(\phi,X)$. One can now easily verify that this action is of the Horndeski type with
\begin{eqnarray}
&&G_2=\hat{K}(\phi,X),\nonumber\\
&&G_3=0,\nonumber\\
&&G_4=\hat{g}(\phi),\nonumber\\
&&G_5=-\hat{h}(\phi)\phi.
\end{eqnarray}
We will not explore more general models here but we will stress that for any choice of non-minimal couplings, i.e., given $f_1(T)$ and $f_3(T)$, the conditions expressed in (\ref{Kessencecondition1}) and  (\ref{Kessencecondition}) will impose tight constraints on the possible form of $K(\phi,X)$. Analogously, inverting the argument, given a function $K(\phi,X)$, said conditions will strongly limit the allowed functions $f_1(T)$ and $f_3(T)$. As a final comment, let us remember that our aim here was to obtain theories with second order equations of motion, so our conditions will be sufficient to avoid the Ostrogradski instability, but models free from such instabilities might also exist. Finally, one could also allow for models where the conformal mode can be excited in a healthy way within the context of $K$-essence models by going to the multi-scalar representation of the theory. In such a formulation, analogous equations might be obtained relating the possible forms of the functions $f$ and $K$ to avoid the Ostrogradski instability for the conformal mode.

\section{Vector fields}
\label{Section:Maxwell}
In the precedent sections we have considered a scalar field as our matter lagrangian. Let us now consider the case of a massive vector field whose lagrangian is given by
\begin{eqnarray}
\mathcal{L}_{m}\,=\, -\frac{1}{4}F_{\mu\nu}F^{\mu\nu}+\frac{1}{2}M^2 A^2,
\label{Lm vector field}
\end{eqnarray}
with $F_{\mu\nu}\equiv \partial_{\mu}\,A_{\nu}- \partial_{\nu}\,A_{\mu}$, and $M$ the mass of the field. Consequently the energy-momentum tensor reads
\begin{eqnarray}
T_{\mu\nu}\,=\,-F_{\mu\alpha}F_{\nu}{}^{\alpha}+\frac{1}{4}g_{\mu\nu}F^2-\frac{M^2}{2}g_{\mu\nu}A^2+M^2 A_{\mu}A_{\nu}\,.\nonumber\\
\label{E-M tensor vector field}
\end{eqnarray}
whose trace and contraction with the Ricci scalar become respectively
\begin{eqnarray}
T\,&=&\,-M^2A^2,\\
R_{\mu\nu}T^{\mu\nu}\,&=&\,\frac{1}{4}\Big(R F_{\mu\nu}F^{\mu\nu}-4R_{\mu\nu}F^{\mu\alpha}F^{\nu}{}_{\alpha}\Big)\nonumber\\%+\frac{1}{4}R\,F^2
 &+& M^2 G_{\mu\nu}A^{\mu}A^{\nu}
%Replaced by previous line -\frac{M^2}{2}R A^2+M^2 R_{\mu\nu}A^{\mu}A^{\nu}
\,.
\label{Contraction vector field}
\end{eqnarray}
As expected, the trace $T$  vanishes for the case of a massless vector field as it corresponds to a conformally invariant theory.  Along the very same line of reasoning as discussed in the previous section for the scalar field case, these theories will typically contain the Ostrogradski instability arising from the non-minimal coupling of the vector field with the scalar curvature. 

In principle one could also attempt to obtain the corresponding Hordenski vector-tensor interaction. In this case however the allowed interaction between the Faraday tensor $F_{\mu\nu}$ and the Ricci curvature 
are much more restricted and in fact there is a unique 
coupling between them that leads to second-order equations of motion. Such a coupling was already computed by Horndeski \cite{Horndeski:1976gi} and it is given by 
\begin{eqnarray}
L^{\alpha\beta\gamma\delta}F_{\alpha\beta}F_{\gamma\delta}\,&=&\,R F_{\mu\nu}F^{\mu\nu} - 4R_{\mu\nu}F^{\mu\sigma}F^{\nu}_{\;\;\sigma} \nonumber\\
&+& R_{\mu\nu\alpha\beta} F^{\mu\nu} F^{\alpha\beta} 
\label{L definition}
\end{eqnarray}
\\
where $L^{\alpha\beta\gamma\delta}=-\frac{1}{2}\epsilon^{\alpha\beta\mu\nu} \epsilon^{\gamma\delta\rho\sigma} R_{\mu\nu\rho\delta}$
stands for the double dual Riemann tensor. The cosmology of this type of interaction was studied in \cite{Barrow:2012ay} and its stability was thoroughly analyzed in \cite{Jimenez:2013qsa}. A careful comparison between
(\ref{Contraction vector field}) and (\ref{L definition}) shows that only the first two terms in the right-hand side %r.h.s. 
of (\ref{L definition}) appear in  (\ref{Contraction vector field}) but the last one in (\ref{L definition}) is absent. Thus one is led to infer that for both massive and massless vector fields, actions of the form
(\ref{Action_fRT})
cannot lead to second-order equations of motion. It should be noticed that kinetic interactions involving second derivatives of the vector field as those considered in \cite{Heisenberg:2014rta} cannot appear because of the gauge invariance of the kinetic term in the Proca action.

%\subsubsection{Multi-scalar representation}
%\label{sub subsection:Multiscalar-tensor representation analysis Vector}

Let us also show for completeness the multi-scalar representation for the vector field just studied. After performing the redefinitions introduced in Section \ref{Section:formalism} the action (\ref{Action_fRT_scalar-tensor representation final}) becomes,
\begin{widetext}
\begin{eqnarray}
S&=&\int {\rm d}^4x\sqrt{-\tilde{g}}\left\{
\e^{4\Omega}\hat{\V}-\frac{1}{16\pi G}\left[\tilde{R}-6(\partial\Omega)^2\right]-\varphi_3M^2\tilde{G}^{\mu\nu}A_{\mu}A_{\nu}
-\varphi_3\e^{-2\Omega} \left(\frac{1}{4}\tilde{F}^2\tilde{R}-\tilde{R}_{\mu\nu}\tilde{F}^{^{\mu\alpha}}\tilde{F}^{\nu}{}_{\alpha}\right)
\right.\nonumber\\&+&
2\,\varphi_3\left[\e^{-2\Omega}\left(\frac14 \tilde{F}^2\gt^{\mu\nu}-\tilde{F}^{\mu\alpha}\tilde{F}^\nu{}_\alpha\right)-M^2\left(\At^2\gt^{\mu\nu}-\At^{\mu}\At^{\nu}\right)\right]\nablat_\mu\nablat_\nu\Omega\nonumber\\&-&\left.
2\,\varphi_3\left[\e^{-2\Omega}\left(\frac14 \tilde{F}^2\gt^{\mu\nu}-\tilde{F}^{\mu\alpha}\tilde{F}^\nu{}_\alpha\right)+M^2\left(\At^2\gt^{\mu\nu}+\At^{\mu}\At^{\nu}\right)\right]\nablat_\mu\Omega\nablat_\nu\Omega
 +\e^{4\Omega}\Lag_m\Big(\e^{2\Omega}\gt_{\mu\nu},\At\Big)\right\}\,,
\label{action_Vector_Field}
\end{eqnarray}
\end{widetext}
where $\tilde{F}^2\equiv \tilde{F}^{\mu\nu}F_{\mu\nu}$, $\tilde{F}^{\mu\nu}\equiv \tilde{g}^{\mu\alpha}\tilde{g}^{\nu\beta}F_{\alpha\beta}$ and $\tilde{F}^{\mu}_{\;\;\nu}\equiv \tilde{g}^{\mu\alpha}F_{\alpha\nu}$. 
Here again, one can identify %the non-Horndeskian couplings between the Ricci tensor and $F^{\mu\nu}$ that will yield 
the instability problems discussed above. Again, as it happened for the scalar field case, the fact that $\varphi_3$ is an auxiliary field will typically lead to undesired terms even if the structure of the couplings of the vector field, curvature and the conformal mode are {\it appropriate}.  However, for theories linear in $R_{\mu\nu}T^{\mu\nu}$, $\varphi_3$ appearing in (\ref{action_Vector_Field}) will not be an auxiliary field, but a simple constant parameter. We will assume this in the following discussion.

The first problem with the vector field could have arisen from the direct coupling of $A_\mu$ to the scalar curvature since such couplings usually introduce additional modes associated with an extra propagating degree of freedom for the vector field. However, for the simple case of a Proca field, the direct coupling to the curvature happens through the Einstein tensor which guarantees the absence of such an extra mode. This can easily be seen by resorting to the Stueckelberg trick, i.e., if we consider the purely longitudinal mode $A_\mu=\partial_\mu\theta$. Such a mode will couple to the Einstein tensor precisely in the form required to be of the Horndeski form and, therefore, it will not introduce any additional modes. This kind of interaction was studied in \cite{BeltranJimenez:2010uh} as a non-minimal coupling to the electromagnetic field that could serve as a mechanism to generate magnetic fields from neutral but rotating bodies. It also arises in a natural manner within the context of Weyl geometries \cite{Jimenez:2014rna}.

Another problem with the above action is the derivative non-minimal coupling, i.e., the coupling between $F_{\mu\nu}$ and the curvature. Again, this is not of the Horndeski vector-tensor type of interaction and, thus, it will lead to higher-order equations of motion with the Ostrogradski instability.  This would be enough to prove the instability of these theories in curved backgrounds. However, also the coupling of the conformal mode to $F_{\mu\nu}$ is pathological. Such a coupling possesses a structure of the form $\e^{2\Omega}\left(F^{\mu\alpha}F^\nu{}_\alpha-\frac14 F^2\gt^{\mu\nu}\right)\nablat_{\mu}\nablat_{\nu}\Omega$. If we focus on second time derivatives, we find $(1/2F_{0i}F^{0i}-1/4F_{ij}F^{ij})\nablat^{0}\nablat_{0}\Omega$ that will lead to higher-order equations of motion due to the presence of the $F_{0i} F^{0i}$ term. 

We can conclude that gravitational theories described by $f(R,T,R_{\mu\nu}T^{\mu\nu})$ lead to Ostrogradski instabilities in a very general manner when coupled to vector fields. We therefore find it reasonable to abandon the universality of the non-minimal couplings of these theories and consider them only for specific forms of the matter sector, e.g. a certain class of scalar fields, for which the instabilities can be avoided.

%%%%%%%%%%%%%%%%%%%%%%%%%%%%%%

\section{Particular models}
\label{Section:Particular models}
In this section we will show how our general discussions and results for the gravitational theories under study in this work for specific realizations. We will choose the models so we can explicitly see some of the points raised above. For simplicity, in the following we will consider a canonical scalar field as corresponding to the matter Lagrangian.

%The purpose of this section is to analyze some models which are studied at the moment, however they would be unstable since the Ostrogradski instability. This type of instabilities introduce causes like the system can decay to positive or negative values infinitely in our theory \cite{Woody}, so this instability would do non-viable our model and we must to discard it.
%
%Our procedure, briefly explain in the previous sections, consists in analyze the action associated with the model which we want to analyze and look for terms of Horndesky type for avoid this instability.
%
%Therefore this section is organized in five subsection where we analyze five models studied like possible theories for modified gravity \cite{Haghani:2013oma}. At the end of this section, we can say if the model if agreement with Horndesky theorem or if the model needs some condition.

%
%\vspace{0.5cm}
\subsection{ Model I: $ f(R,T,R_{\mu\nu}T^{\mu\nu})=\alpha R^n+\beta(R_{\mu\nu}T^{\mu\nu})^m$}

\label{subsection Rn RTm }
We start by considering a superposition of two power laws so that the action takes the form
\begin{eqnarray}
S_{\rm I}&=&\int\diff^4x\sqrt{-g}\left[ \alpha R^n+\beta \Big(G^{\mu\nu}\partial_\mu\phi\partial_\nu\phi+RV(\phi)\Big)^m\right.\nonumber\\
&-&\left.\frac12(\partial \phi)^2-V(\phi)\right].
\end{eqnarray}
The definitions for the auxiliary fields $\varphi_i=-\frac{\partial f}{\partial \chi_i} $ in this case yield $\varphi_1=-n \alpha \chi_1^{n-1}$ and $\varphi_3=-m\beta \chi_3^{m-1}$. We see that the condition for the invertibility of these relations is that $m$ and $n$ are different from 1, i.e., whenever the dependence on $R$ and $R_{\mu\nu} T^{\mu\nu}$ is not linear. In the case of $n=m=1$ the theory is stable because we simply have the Einstein-Hilbert term plus a canonical scalar field with a non-minimal derivative coupling to the Einstein tensor which, as discussed several times throughout this work, corresponds to a healthy coupling. Moreover, even for $n\neq1$ the theory will be free of Ostrogradski instabilities. To see this more clearly it is convenient to go to the multi-scalar representation. After the appropriate conformal transformation, our action for arbitrary $n$ and $m=1$ reads:
%%Now, we propose the following conformal transformation:
%%\begin{eqnarray}
%%\tilde{g}_{\mu\nu}=e^{-2\Omega}g_{\mu\nu},
%%\end{eqnarray}
%
%where:
%\begin{eqnarray}
%\Omega=log{\frac{1}{\sqrt{16\pi G[\varphi_1+\varphi_3 V(\phi)]}}},
%\end{eqnarray}
\begin{eqnarray}
S_{\rm I}&=&\int {\rm d}^4x\sqrt{-\tilde{g}}\Big\{ \e^{4\Omega}\mathcal{U}(\Omega)-\frac{1}{16\pi G}\left[\tilde{R}-6(\partial\Omega)^2\right]\nonumber\\
&-&\beta\left[\tilde{G}_{\mu\nu}\partial^{\mu}\phi\partial^{\nu}\phi+2(\partial_{\alpha}\Omega\partial^{\alpha}\Omega)^2+(\partial\Omega)^2(\partial\phi)^2
\right.\nonumber\\
&+&\left.2\left(\tilde{g}^{\mu\nu}(\partial\phi)^2-(\partial^\mu\phi\partial^\nu\phi)\right)\tilde{\nabla}_\mu\tilde{\nabla}_\nu\Omega\right]
\nonumber\\
&+&\e^{4\Omega}\mathcal{L}_{m}(\phi,{\e}^{2\Omega}\tilde{g}_{\mu\nu})\Big\}.%\nonumber\\
\end{eqnarray}
Here again we see that the conformal mode couples to the scalar field in the appropriate way not to lead to higher-order equations of motion and the derivative non-minimal coupling of the scalar field $\phi$ belongs to the Horndeski type. Therefore, for this case the theory avoids the Ostrogradski instability. However, for arbitrary $m$, the coupling $R_{\mu\nu}T^{\mu\nu}$ will spoil the nice structure of the above equation and, therefore, the Ostrogradski instability will reappear. A particular case with $n=1$ and arbitrary $m$ is a special case that we treat in more detail in the next subsection.

%With the conformal transformation, we have achieved to obtains the Einstein-Hilbert action (the Ricci's scalar is uncoupling) and we have eliminated the potential $V(\phi)$
%
%It easy to see too, that due the presence of $\varphi_3$, like a dynamical field, this action will present Ostrogradski's instabilities in general, so that, although the term between brackets don't introduce instabilities (we have used the Appendix-A for this conclusion), $\varphi_3$ could introduce more degrees of freedom.

%It's very important to note that when we have done the Legendre transformation, we had imposed that $\chi_3=R_{\mu\nu}T^{\mu\nu}$; so the main conclusion is that pur theory must be constant for the coupling between R and T.

%If $n=1$ or $m=1$ we can't do the Legendre's transformation but, we don't necessitate do it because our action would be lineal in $R$ and $R_{\mu\nu}T^{\mu\nu}$. This particular case will be studied more forward.

%
\vspace{0.5cm}
\subsection{Model II:  $f(R,T,R_{\mu\nu}T^{\mu\nu})=-\frac{R}{16\pi G}+\beta(R_{\mu\nu}T^{\mu\nu})^m$}
\label{R+RTm}

This is a particular case of the class of models above with $n=1$. This case is special because the auxiliary field $\varphi_1$ which is usually mapped into the conformal mode cannot be defined due to the non-invertibility of the Legendre transformation for $\chi_1$. Thus, we can only introduce the field $\chi_3$ and it is related to $\varphi_3$ by $-\varphi_3=-m\beta\chi_3^{m-1}$. Again, for $m=1$ this transformation is not invertible and must be treated separately (see the end of this section). The action in terms of the auxiliary fields now yields
%\begin{eqnarray}
%S&=&\int {\rm d}^4x\sqrt{-{g}}\left\{\matcal-\frac{R}{16\pi G}+\beta(G^{\mu\nu}\partial_\mu\phi\partial_\nu\phi+RV(\phi)\Big)^m+\mathcal{L}_{m}(\phi,g_{\mu\nu})\right\}
%\end{eqnarray}
\begin{eqnarray}
S_{\rm II}&=&\int {\rm d}^4x\sqrt{-{g}}\left[\mathcal{U}(\varphi_3)-\left(\frac{1}{16\pi G}+\varphi_3 V(\phi)\right)R\right.\nonumber\\&-&\left.\varphi_3 G^{\mu\nu}\partial_\mu\phi\partial_\nu\phi+\mathcal{L}_{m}(\phi,g_{\mu\nu})\right],
\end{eqnarray}
with $\mathcal{U}(\varphi_3)=\beta(1-m)\left(\frac{-\varphi_3}{m\beta}\right)^{\frac{m}{m-1}}$. Accordingly we can see that the conformal mode will still be excited but now it will be originated from the auxiliary field $\varphi_3$. Therefore, the auxiliary field $\varphi_3$ will disappear from the action after performing the conformal transformation. In fact, the required conformal factor is now given by $\varphi_3=\frac{\e^{-2\Omega}-1}{16\pi G V(\phi)}$ and the resulting action reads
%
%We can see we have the Einstein-Hilbert action plus the coupling between R and T raised to the m-th power.
%The definitions for the auxiliary field in this case yield $\varphi_3=-m \beta\chi_3^{m-1}$

%Then, we introduce the Legendre transformation again:
%\begin{eqnarray}
%S&=&\int {\rm d}^4x\sqrt{-{g}}\Big\{\mathcal{U}(\varphi_3)-\left(\frac{1}{16\pi G}+\varphi_3 V(\phi)\right)R-\varphi_3 G^{\mu\nu}\partial_\mu\phi\partial_\nu\phi+\mathcal{L}_{m}(\phi,g_{\mu\nu})\Big\},
%\end{eqnarray}

%where: $\varphi_3=-m\beta^{1-m}\chi_3^{m-1}$.Now, we conformal transformation would can be:
%\begin{eqnarray}
%\tilde{g}_{\mu\nu}=e^{-2\Omega}g_{\mu\nu},
%\end{eqnarray}

%where:
%\begin{eqnarray}
%\Omega=log{\frac{1}{\sqrt{16\pi G[\frac{1}{16\pi G}+\varphi_3V(\phi)]}}}
%\end{eqnarray}
\begin{eqnarray}
S_{\rm II}&=&\int {\rm d}^4x\sqrt{-\tilde{g}}\Big\{\e^{4\Omega}\mathcal{U}(\Omega,\phi)-\frac{1}{16\pi G}\Big(\tilde{R}+6(\partial\Omega)^2\Big)\nonumber\\
&+&\frac{1-{\rm e}^{-2\Omega}}{16\pi G V(\phi)}\Big[\tilde{G}^{\mu\nu}\partial_\mu\phi\partial_\nu\phi+2(\partial^\alpha\Omega\partial_\alpha\phi)^2\nonumber\\
&+&(\partial\Omega)^2(\partial\phi)^2+2(\tilde{g}^{\mu\nu}(\partial\phi)^2-\partial^{\mu}\phi\partial^\nu\phi)\tilde{\nabla}_\mu\tilde{\nabla}_\nu\Omega\Big]\nonumber\\
&+&{\rm e}^{4\Omega}\mathcal{L}_{m}(\phi,e^{2\Omega}\tilde{g}_{\mu\nu})\Big\},
\end{eqnarray}
where we can see that all the interactions containing second order derivatives have the appropriate form to avoid higher-order equations of motion. Thus, even though the starting action is not linear in $R_{\mu\nu}T^{\mu\nu}$, the fact that it is linear in the scalar curvature makes the theory free from Ostrogradski instabilities. The underlying reason, as we have shown, is that one cannot define the auxiliary field $\varphi_1$ and this actually ensures that it is the auxiliary field $\varphi_3$ the one that will be mapped into the conformal mode and, therefore, it will not be present in the final action after the conformal transformation.

%Again, the Einstein's tensor is coupling with $\varphi_3$ which is a dynamic field so that it can introduce more degree of freedom in the action and then the action will have got instabilities of Ostrogradski type, i.e, we have obtained the same result to the last case. Therefore, both theories are the same for Ostrogradski instabilities. This result was expected because f(R) theories don't have this instabilities and so the term $R^n$ in the previous section don't introduce problems.

%Again, we must to impose $\chi_3=constant$ with the purpose of it doesn't introduce derivatives term, i.e, we have obtained the same condition to the last case. Therefore, both theories are the same for Ostrogradski instabilities. This result was expected because f(R) theories don't have this instabilities and so the term $R^n$ in the previous section don't introduce problems.

%

%\vspace{0.5cm}
%\subsection{Model III: $f(R,T,R_{\mu\nu}T^{\mu\nu})=-\frac{1}{16\pi G}R+\beta R_{\mu\nu}T^{\mu\nu}$}

%Now, the action takes the form:
%\begin{eqnarray}
%S&=&\int {\rm d}^4x\sqrt{-{g}}\Big\{R+\alpha R_{\mu\nu}T^{\mu\nu} +\mathcal{L}_{m}(\phi,g_{\mu\nu})\Big\}
%\end{eqnarray}

The case with $m=1$ was studied in \cite{Haghani:2013oma} and corresponds to the particular model described by the action
\begin{eqnarray}
S_{\rm II}&=&\int {\rm d}^4x\sqrt{-{g}}\Big\{\left[1+\alpha V(\phi)\right]R+\alpha G^{\mu\nu}\partial_\mu\phi\partial_\nu\phi\nonumber\\
&+&\mathcal{L}_{m}(\phi,g_{\mu\nu})\Big\}
\end{eqnarray}
where we see that the theory belongs to the Horndeski class for the scalar field $\phi$ and the conformal mode is not excited.
\vspace{0.5cm}
\subsection{Model III: $f(R,T,R_{\mu\nu}T^{\mu\nu})=\alpha R\left(1+\beta R_{\mu\nu}T^{\mu\nu}\right)$ case}
%Then:
%\bea
%S&=&\int {\rm d}^4x\sqrt{-{g}}\Big\{R\left[1+\alpha R_{\mu\nu}T^{\mu\nu}\right]+\mathcal{L}_{m}(\phi,g_{\mu\nu})\Big\}
%\eea
%
Now we will consider a slightly more involved model where the different arguments of the function $f$ appear mixed. The field redefinition from the fields $\chi_i$ to the fields $\varphi_i$ is given (notice that we only have $\chi_1$ and $\chi_3$ but not $\chi_2$ because there is no dependence on $T$ in the action): $\varphi_1=-\alpha(1+\beta\chi_3)$ and $\varphi_3=-\alpha\beta\chi_1$. This field redefinition is actually an injective transformation for $\alpha\neq0$ and $\beta\neq0$ so that it is perfectly valid. Moreover, notice that the matrix $\partial^{2} f/\partial\chi_i\partial\chi_j$ has determinant $\det(\partial^{2} f/\partial\chi_i\partial\chi_j)=-(\alpha\beta)^2$ which is non-vanishing and, therefore, the Legendre transformation is legitimate. After introducing the auxiliary fields $\varphi_i$ the action reads
\bea
S_{\rm III}&=&\int\diff^4x\sqrt{-g}\left[\mathcal{U}(\varphi_1,\varphi_3)-\Big(\varphi_1+\varphi_3 V(\phi)\Big)R\right.\nonumber\\
&-&\left.\varphi_3G^{\mu\nu}\partial_\mu\phi\partial_\nu\phi\right]
\eea
with $\mathcal{U}(\varphi_1,\varphi_3)=\varphi_3\chi_3=-\frac1\beta\varphi_3(1+\frac1\alpha\varphi_1)$. We see in this case that both $\varphi_1$ and $\varphi_3$ appear linearly so they could be seen as Lagrange multiplier fields. However, since $\partial^2\mathcal{U}/\partial\varphi_i\partial\varphi_j$ is a non-degenerate matrix (which coincides with the inverse of $-\partial^2 f/\partial\chi_i\partial\chi_j$, as we explained above), the equations of motion for $\varphi_1$ and $\varphi_3$ actually allow us to algebraically solve them, so they are indeed auxiliary fields. This will be the main obstruction for the stability of this theory because that will make the term $G^{\mu\nu}\partial_\mu\phi\partial_\nu\phi$ appear non-linearly in the action and, therefore, the Ostrogradski instability will be present. Notice also that the conformal mode will be excited and can be associated with $\varphi_1$ so that we do not encounter the situation that allowed us to assure the stability of the model II where the field $\varphi_3$ could be used to excite the conformal mode and, therefore, it disappeared from the transformed action.
%
%This action is very different to the previous action because, now we have coupling between $R$ and $R_{\mu\nu}T^{\mu\nu}$. Besides $R$ and $R_{\mu\nu}T^{\mu\nu}$ are linear so, we can't to apply a Legendre transformation. Therefore, we go to develop the action:
%\bea
%S&=&\int {\rm d}^4x\sqrt{-{g}}\Big\{R\left[1+\alpha G^{\mu\nu}\partial_\mu\phi\partial_\nu\phi\right]+R^2 V(\phi)+\mathcal{L}_{m}(\phi,g_{\mu\nu})\Big\}
%\eea
%
%Now, we can apply a Legendre transformation for the term $R^2 V(\phi)$ and therefore:
%\bea
%S=\int {\rm d}^4x\sqrt{-{g}}\Big\{\mathcal{U}(\varphi_1,\phi)+R\left[1+\alpha G^{\mu\nu}\partial_\mu\phi\partial_\nu\phi-\varphi_1 V(\phi)\right]+\mathcal{L}_{m}(\phi,g_{\mu\nu})\Big\}
%\eea
%
%%where in this case: $\mathcal{U}(\varphi_1,phi)=\chi_1^2 V(\phi)+\varphi_1\chi_1 V(\phi)$ and $\varphi_1=-2\chi_1$
%where in this case: $\mathcal{U}(\varphi_1,\phi)=\frac{-1}{4}\varphi_1^2 V(\phi)$ and $\varphi_1=-2\chi_1$
%
%
%But this is horrible because is totally different to Horndeski action so we can't say this action is ghosts-free and thus we don't accept this model in general.
%

%%%%%%%%%%%%%%%%%%%%%%%%%%%%%%

\section{Conclusions}
\label{Section:Conclusions}

In this work we have considered a class of universal non-minimally coupled theories of gravity where the non-minimal coupling is achieved through couplings of the energy-momentum tensor of the matter sector to the curvature, i.e.,  the gravitational Lagrangian is of the form $f(R,T,R_{\mu\nu} T^{\mu\nu})$. These theories have received some attention in recent literature and might offer interesting cosmological applications. The aim of this work has been to clarify some issues concerning the consistency and stability of such theories. First of all, we discussed the fact that the energy-momentum tensor appears at the level of the action and this might pose consistency problems since the energy-momentum tensor is a quantity to be derived from the action. Another way of expressing the potential inconsistency is that the energy-momentum tensor is the conserved current associated with infinitesimal translations. However, if plugged back in the action this statement is no longer true and, in fact, such a quantity will not be conserved anymore. As we argued in section \ref{Section:formalism}, for the considered theories, the object entering the action is not really the conserved energy-momentum tensor and should be regarded simply as an operational manner to define the theory.

Secondly, another potential problem with these theories is the presence of second order derivatives in the Lagrangian that will typically lead to higher-order equations of motion with the associated Ostrogradski instability. The study of this issue has comprised the majority of this work. We have first shown the problem for a general matter Lagrangian. Moreover, we have also allowed for cases where the conformal mode can also be excited. We have found two sources from which the Ostrogradski might appear. The first one is the generation of derivative non-minimal couplings of the matter fields to curvature, while the second one corresponds to the conformal mode entering with second derivatives in the action. After presenting the general framework to analyze the presence of Ostrogradski instabilities for these types of theories we have considered specific cases for the matter sector, namely: a canonical scalar field, a $K$-essence field and, finally, a vector field. For the canonical scalar field we have found conditions under which both the scalar field and the conformal mode lead to second order equations of motion. For that we have imposed that the corresponding terms in the action are of the Horndeski form, as those are the most general terms for scalar-tensor theories yielding second order equations of motion. However, as we have already commented, those do not correspond to the most general scalar-tensor theory free from the Ostrogradski instability so that our results can be viewed as sufficient conditions, although more general theories might still be possible. Subsequently to the detailed analysis for the canonical scalar field case we have considered general $K$-essence models. In these theories, the presence of one extra free function opens new possibilities. Thus we have obtained the conditions to avoid higher-order equations of motion for general $K$-essence models. 

Finally, after studying the case of scalar fields, we have considered the case of a Proca vector field. Unlike the scalar field case, we have found that it is not possible to obtain viable models free from the Ostrogradski instability. This led us to conclude that the universal nature of the non-minimal coupling should be abandoned because, although it is possible to obtain stable models for scalar fields, it is troublesome to have couplings to vector fields.

%%%%%%%%%%%%%%%%%%%%%%%%%%%%%%%%%%%%%%%%%
\begin{acknowledgements}
We would like to thank Tomi S. Koivisto for useful comments.
I.A. has received financial support from the Consolider-Ingenio MULTIDARK CSD2009-00064 Summer 2014 grants.
J.B.J. is supported by the Wallonia-Brussels Federation grant ARC No. 11/15-040 and also acknowledges the financial support of A*MIDEX project (ANR-11-IDEX-0001-02) funded by the "Investissements d$'$Avenir" French Government program, managed by the French National Research Agency (ANR)
A.d.l.C.D. acknowledges financial support from MINECO (Spain) project FPA2011-27853-C02-01. 
J.B.J. and A.d.l.C.D. acknowledge financial support from MINECO (Spain) projects FIS2011-23000 and Consolider-Ingenio MULTIDARK CSD2009-00064.
I.A. thanks the Theoretical Physics I Department (Madrid U.) for the use of facilities.
A.d.l.C.D. is indebted to the Centre de Cosmologie, Physique des Particules et Ph\'enom\'enologie CP3, 
Universit\'e catholique de Louvain, Louvain-la-Neuve, Belgium for its assistance with the first steps of this manuscript and the 
Astrophysics, Cosmology and Gravity Centre (ACGC), Department of Mathematics and Applied Mathematics, University of Cape Town, South Africa during the latest stages of its preparation. 
\end{acknowledgements}

%%%%%%%%%%%%%%%%%%%%%%%%%%%%%%%%%%%%%%%%%


\begin{thebibliography}{99} 




 \bibitem{Scalar-tensor proposals}
   %C. Brans and R. H. Dicke, Phys. Rev., {\bf 124} 925 (1961);
  C. H. Brans, Phys. Rev., {\bf 125(6)} 2194 (1962);
  J. Garc\'ia-Bellido, A. Linde, and D. Linde, Phys. Rev. D, {\bf 50} 730 (1994);
  J.~A.~R.~Cembranos  {\it et al.}, JCAP {\bf 0907}, 025 (2009) [0905.1989 [astro-ph.CO]]; 
  T.~Biswas {\it et al.}, Phys.\ Rev.\ Lett.\  {\bf 104}, 021601 (2010) [arXiv:0910.2274 [hep-th]]; 
  JHEP {\bf 1010}, 048 (2010) [arXiv:1005.0430 [hep-th]]; 
  Phys.\ Rev.\  D {\bf 82}, 085028 (2010) [arXiv:1006.4098 [hep-th]]  
  
  
 
 \bibitem{Vector proposals}
  L.~H.~Ford, Phys.\ Rev.\ D {\bf 40} (1989) 967; 
  T.~Koivisto and D.~F.~Mota, JCAP {\bf 0808}, 021 (2008) [arXiv:0805.4229 [astro-ph]]; %%CITATION = ARXIV:0805.4229;%%
  J.~A.~R.~Cembranos  {\it et al.}, Phys.\ Rev.\ D {\bf 86}, 021301 (2012); [arXiv:1203.6221 [astro-ph.CO]];   
  	Phys. Rev. D {\bf 87}, 043523 (2013)  arXiv:1212.3201 [astro-ph.CO];
	%\cite{Jimenez:2009py}
%\bibitem{Jimenez:2009py}
  J.~Beltran Jimenez, R.~Lazkoz and A.~L.~Maroto,
  %``Cosmic vector for dark energy: Constraints from supernovae, cosmic microwave background, and baryon acoustic oscillations,''
  Phys.\ Rev.\ D {\bf 80} (2009) 023004
  [arXiv:0904.0433 [astro-ph.CO]];
  %%CITATION = ARXIV:0904.0433;%%
  %23 citations counted in INSPIRE as of 06 Nov 2014 
%\cite{Jimenez:2008au}
  J.~B. Jim\'enez and A.~L.~Maroto,  %``A cosmic vector for dark energy,''
  Phys.\ Rev.\ D {\bf 78} (2008) 063005   [arXiv:0801.1486 [astro-ph]];
  %%CITATION = ARXIV:1205.1695;%%
  %7 citations counted in INSPIRE as of 06 Nov 2014
%\cite{Jimenez:2008nm}
%  J.~Beltran Jimenez and A.~L.~Maroto,
  %``Cosmological electromagnetic fields and dark energy,''
  JCAP {\bf 0903} (2009) 016  [arXiv:0811.0566 [astro-ph]];
  %\cite{Jimenez:2009sv}
%\bibitem{Jimenez:2009sv}
  %J.~Beltran Jimenez, T.~S.~Koivisto, A.~L.~Maroto and D.~F.~Mota,
  %``Perturbations in electromagnetic dark energy,''
  JCAP {\bf 0910} (2009) 029
  [arXiv:0907.3648 [physics.gen-ph]];
  %%CITATION = ARXIV:0907.3648;%%
  %27 citations counted in INSPIRE as of 06 Nov 2014
%\cite{Jimenez:2008sq}
%  J.~Beltran Jimenez and A.~L.~Maroto,   %``Viability of vector-tensor theories of gravity,''
  JCAP {\bf 0902} (2009) 025   [arXiv:0811.0784 [astro-ph]];
%\cite{Jimenez:2009dt}
 % J.~Beltran Jimenez and A.~L.~Maroto,   %``The electromagnetic dark sector,''
  Phys.\ Lett.\ B {\bf 686} (2010) 175   [arXiv:0903.4672 [astro-ph.CO]];
  %\cite{Jimenez:2010hu}
%\bibitem{Jimenez:2010hu}
  %J.~Beltran Jimenez and A.~L.~Maroto,
  %``Cosmological magnetic fields from inflation in extended electromagnetism,''
  Phys.\ Rev.\ D {\bf 83} (2011) 023514
  [arXiv:1010.3960 [astro-ph.CO]];
  %%CITATION = ARXIV:1010.3960;%%
  %11 citations counted in INSPIRE as of 06 Nov 2014
%\cite{Jimenez:2009ai}
 %  J.~Beltran Jimenez and A.~L.~Maroto,   %``Cosmological evolution in vector-tensor theories of gravity,''
  Phys.\ Rev.\ D {\bf 80} (2009) 063512   [arXiv:0905.1245 [astro-ph.CO]];
    %\cite{Carlesi:2012rq}
%\bibitem{Carlesi:2012rq}
  E.~Carlesi, A.~Knebe, G.~Yepes, S.~Gottloeber, J.~Beltran Jimenez and A.~L.~Maroto,
  %``N-body simulations with a cosmic vector for dark energy,''
  Mon.\ Not.\ Roy.\ Astron.\ Soc.\  {\bf 424} (2012) 699
  [arXiv:1205.1695 [astro-ph.CO]];
  %\cite{Carlesi:2011gv}
%\bibitem{Carlesi:2011gv}
  %E.~Carlesi, A.~Knebe, G.~Yepes, S.~Gottloeber, J.~Beltran Jimenez and A.~L.~Maroto,
  %``Vector dark energy and high-z massive clusters,''
  Mon.\ Not.\ Roy.\ Astron.\ Soc.\  {\bf 418} (2012) 2715
  [arXiv:1108.4173 [astro-ph.CO]];
  %%CITATION = ARXIV:1108.4173;%%
  %13 citations counted in INSPIRE as of 06 Nov 2014
%\cite{BeltranJimenez:2013fca}
  J.~Beltran~Jimenez, A.~L.~Delvas Froes and D.~F.~Mota,   %``Screening Vector Field Modifications of General Relativity,''
  Phys.\ Lett.\ B {\bf 725} (2013) 212   [arXiv:1212.1923 [astro-ph.CO]].
  
%  \cite{Clifton:2011jh}
%T.~Clifton, P.~G.~Ferreira, A.~Padilla and C.~Skordis,   %``Modified Gravity and Cosmology,''
% Phys.\ Rept.\  {\bf 513} (2012) 1   [arXiv:1106.2476 [astro-ph.CO]].
  
\bibitem{matter-geometry couplings}
S. Nojiri and S. D. Odintsov, % "Gravity assisted dark energy dominance and cosmic acceleration"
Phys. Lett. B {\bf 599}, 137 (2004); 
%
G. Allemandi, A. Borowiec, M. Francaviglia, S. D. Odintsov, % Dark energy dominance and cosmic acceleration in first order formalism
Phys. Rev. D {\bf 72} 063505 (2005) [arXiv:0504057 [gr-qc]];
% T. Koivisto, Class. Quant. Grav. 23, 4289 (2006).
V. Faraoni, Phys. Rev. D 76, 127501 (2007); Phys. Rev. D {\bf 80}, 124040 (2009); 
T. P. Sotiriou, % "The viability of theories with matter coupled to the Ricci scalar"
Phys. Lett. B 664, 225 (2008); 
O. Bertolami, F. S. N. Lobo, and J. Paramos, Phys. Rev. D {\bf 78}, 064036 (2008); 
S. Nesseris,  %Matter density perturbations in modified gravity models with arbitrary coupling between matter and geometry"
Phys. Rev. D {\bf 79}, 044015 (2009) arXiv:0811.4292 [astro-ph];
T. Harko, Phys. Rev. D {\bf 81}, 084050 (2010); Phys. Lett. B 669, 376 (2008); 
%\cite{Harko:2010hw}
  T.~Harko, T.~S.~Koivisto and F.~S.~N.~Lobo,   %``Palatini formulation of modified gravity with a nonminimal curvature-matter coupling,''
  Mod.\ Phys.\ Lett.\ A {\bf 26}, 1467 (2011)   [arXiv:1007.4415 [gr-qc]]; 
 S. Thakur, A. A. Sen, and T. R. Seshadri, Phys. Lett. B 696, 309 (2011); 
%\cite{Mukohyama:2003nw}
  S.~Mukohyama and L.~Randall,   %``A Dynamical approach to the cosmological constant,''
  Phys.\ Rev.\ Lett.\  {\bf 92}, 211302 (2004)   [hep-th/0306108].
%\cite{Deruelle:2008fs}
  N.~Deruelle, M.~Sasaki and Y.~Sendouda,   %``'Detuned' f(R) gravity and dark energy,''
  Phys.\ Rev.\ D {\bf 77}, 124024 (2008)   [arXiv:0803.2742 [gr-qc]].
%\cite{Harko:2012hm}
  T.~Harko, F.~S.~N.~Lobo and O.~Minazzoli,   %``Extended $f(R,L_m)$ gravity with generalized scalar field and kinetic term dependences,''
  Phys.\ Rev.\ D {\bf 87}, no. 4, 047501 (2013)  [arXiv:1210.4218 [gr-qc]];
  %\cite{Heisenberg:2014rka}
%\bibitem{Heisenberg:2014rka}
  L.~Heisenberg,
  %``Quantum corrections in massive bigravity and new effective composite metrics,''
  arXiv:1410.4239 [hep-th];
  %%CITATION = ARXIV:1410.4239;%%
  %1 citations counted in INSPIRE as of 06 Nov 2014
  %\cite{deRham:2014naa}
%\bibitem{deRham:2014naa}
  C.~de Rham, L.~Heisenberg and R.~H.~Ribeiro,
  %``On couplings to matter in massive (bi-)gravity,''
  arXiv:1408.1678 [hep-th];
  %%CITATION = ARXIV:1408.1678;%%
  %21 citations counted in INSPIRE as of 06 Nov 2014
  %\cite{deRham:2014fha}
%\bibitem{deRham:2014fha}
  C.~de Rham, L.~Heisenberg and R.~H.~Ribeiro,
  %``Ghosts & Matter Couplings in Massive (bi-&multi-)Gravity,''
  arXiv:1409.3834 [hep-th].
  %%CITATION = ARXIV:1409.3834;%%
  %14 citations counted in INSPIRE as of 06 Nov 2014
 
 
%\cite{Tamanini:2013aca}
\bibitem{Koivisto 2013} 
  N.~Tamanini and T.~S.~Koivisto,   %``Consistency of nonminimally coupled $f(R)$ gravity,''
  Phys.\ Rev.\ D {\bf 88}, no. 6, 064052 (2013)   [arXiv:1308.3401 [gr-qc]].
  
  
\bibitem{varia-fR-Lagrangian}
O. Bertolami, C. G. Bohmer, T. Harko and F. S. N. Lobo, Phys. Rev. D {\bf 75} (2007) 104016 [arXiv:0704.1733 [gr-qc]];  
T. P. Sotiriou and V. Faraoni, Class. Quant. Grav. {\bf 25} (2008) 205002 [arXiv:0805.1249 [gr-qc]]; 
C. Corda, Mod. Phys. Lett. A {\bf 23} (2008) 109 [arXiv:0801.0319 [astro-ph]]; 
O. Bertolami, J. Paramos, T. Harko and F. S. N. Lobo, contribution to the volume {\it The Problems of Modern Cosmology}, arXiv:0811.2876 [gr-qc]; 
O. Bertolami, P. Frazao and J. Paramos, Phys. Rev. D {\bf 81} (2010) 104046 [arXiv:1003.0850 [gr-qc]];  Phys. Rev. D 83 (2011) 044010 [arXiv:1010.2698 [gr-qc]]; 
D. Puetzfeld and Y. N. Obukhov, Phys. Rev. D {\bf 87} (2013) 044045 [arXiv:1301.4341 [gr-qc]].

\bibitem{DK} A. D. Dolgov and M. Kawasaki, Phys. Lett. B {\bf 573}, 1 (2003).

\bibitem{Sergei_DK} 
S. Nojiri, S. D. Odintsov, %Modified gravity with negative and positive powers of the curvature: Unification of the inflation and of the cosmic acceleration
Phys. Rev. D {\bf 68} (2003) 123512 [arXiv:0307288 [hep-th]].

\bibitem{DK-HuSawicki} W. Hu and I. Sawicki. Phys. Rev. D {\bf 76}, 064004 (2007).

\bibitem{Faraoni2007} V. Faraoni, Phys. Rev. D {\bf 75}, 067302 (2007).

\bibitem{Dolgov-Kawasaki-non-minimal}
J. Wang {\it et al.}, %Conditions and instability in f(R) gravity with non-minimal coupling between matter and geometry
Eur. Phys. J. C {\bf 69}, Numbers 3-4, 541-546, (2010) [arXiv:1212.4928 [gr-qc]];
O. Bertolami and M. C. Sequeira, % Energy Conditions and Stability in f(R) theories of gravity with non-minimal coupling to matter
Phys.Rev. D {\bf 79} 104010 (2009) [arXiv:0903.4540 [gr-qc]]

\bibitem{Odintsov:2013iba}
  S.~D.~Odintsov and D.~S\'aez-G\'omez,   %``$f(R, T, R_{\mu\nu} T^{\mu\nu})$ gravity phenomenology and $\Lambda$CDM universe,''
  Phys.\ Lett.\ B {\bf 725} (2013) 437   [arXiv:1304.5411 [gr-qc]].

\bibitem{Haghani:2013oma}
  Zahra Haghani, T.~Harko, F.~S.~N.~Lobo, H.~R.~Sepangi and S.~Shahidi,   %``Further matters in space-time geometry: f(R,T,R??T??) gravity,''
  Phys.\ Rev.\ D {\bf 88} (2013) 4,  044023   [arXiv:1304.5957 [gr-qc]].
  
  
\bibitem{Ostrogradski} M. Ostrogradski, Mem. Ac. St. Petersbourg VI {\bf 4}, 385 (1850). %  R. Woodard, Lect. Notes Phys. {\bf 720}, 403 (2007) [arXiv:astro-ph/0601672]

\bibitem{Woody} 
  R.~P.~Woodward,   %``Avoiding dark energy with 1/r modifications of gravity,''
  Lect.\ Notes Phys.\  {\bf 720}, 403 (2007)   [astro-ph/0601672].
  

  
%\cite{Jimenez:2012ak}
\bibitem{Jimenez:2012ak}
  J.~B.~Jimenez, E.~Dio and R.~Durrer,   %``A longitudinal gauge degree of freedom and the Pais Uhlenbeck field,''
  JHEP {\bf 1304} (2013) 030
  [arXiv:1211.0441 [hep-th]].
 
  
\bibitem{Horndeski} G. W. Horndeski, Int. J. Theor. Phys. {\bf 10} (1974) 363-384

  \bibitem{Galileons}
A. Nicolis, R. Rattazzi and E. Trincherini, [arXiv:0811.2197 [hep-th]]; 
C. Deffayet, G. Esposito-Farese, A. Vikman, Phys. Rev. D {\bf 79}, 084003 (2009) [arXiv:0901.1314];
%\cite{Deffayet:2011gz}
%\bibitem{Deffayet:2011gz}
  C.~Deffayet, X.~Gao, D.~A.~Steer and G.~Zahariade,
  %``From k-essence to generalised Galileons,''
  Phys.\ Rev.\ D {\bf 84} (2011) 064039
  [arXiv:1103.3260 [hep-th]].
  %%CITATION = ARXIV:1103.3260;%%
  %201 citations counted in INSPIRE as of 18 Nov 2014

\bibitem{Piazza}
%\cite{Gleyzes:2014dya} \bibitem{Gleyzes:2014dya}
  J.~Gleyzes, D.~Langlois, F.~Piazza and F.~Vernizzi,   %``Healthy theories beyond Horndeski,''
  arXiv:1404.6495 [hep-th]; 
  %\bibitem{Gao:2014soa}
  X.~Gao,
  %``Unifying framework for scalar-tensor theories of gravity,''
  Phys.\ Rev.\ D {\bf 90} (2014) 081501
  [arXiv:1406.0822 [gr-qc]].
  %%CITATION = ARXIV:1406.0822;%%
  %12 citations counted in INSPIRE as of 18 Nov 2014
% \bibitem{Lin:2014jga}
  C.~Lin, S.~Mukohyama, R.~Namba and R.~Saitou,  %``Hamiltonian structure of scalar-tensor theories beyond Horndeski,''
  arXiv:1408.0670 [hep-th];
 % \bibitem{Gleyzes:2014qga}
  J.~Gleyzes, D.~Langlois, F.~Piazza and F.~Vernizzi,   %``Exploring gravitational theories beyond Horndeski,''
  arXiv:1408.1952 [astro-ph.CO];
  %\cite{Gao:2014soa}
  %\cite{Gao:2014fra}
%\bibitem{Gao:2014fra}
  X.~Gao,
  %``Hamiltonian analysis of spatially covariant gravity,''
  arXiv:1409.6708 [gr-qc].
  %%CITATION = ARXIV:1409.6708;%%
  %2 citations counted in INSPIRE as of 18 Nov 2014

\bibitem{Zuma} %\cite{Zumalacarregui:2013pma}
M.~Zumalac\'arregui and J.~Garc\'ia-Bellido,   %``Transforming gravity: from derivative couplings to matter to second-order scalar-tensor theories beyond the Horndeski Lagrangian,''
  Phys.\ Rev.\ D {\bf 89} (2014) 064046   [arXiv:1308.4685 [gr-qc]]; 
D. Bettoni and S. Liberati, Phys. Rev. D {\bf 88}, 084020 (2013) [arXiv:1306.6724 [gr-qc]]






\bibitem{MultiScalar} T. Kobayashi, N. Tanahashi, and M. Yamaguchi, Phys. Rev. D {\bf 88}, 083504 (2013) [arXiv:1308.4798 [hep-th]];  A. Padilla and V. Sivanesan, [arXiv:1210.4026 [gr-qc]];  V. Sivanesan, (2013), [arXiv:1307.8081 [gr-qc]].

\bibitem{Non-local} T. Biswas, E. Gerwick, T. Koivisto, and A. Mazumdar, Phys. Rev. Lett. {\bf 108}, 031101 (2012), [arXiv:1110.5249 [gr-qc]].
% T. Biswas, T. Koivisto, A. Mazumdar, % Nonlocal theories of gravity: the flat space propagator
%
Proceedings of the Barcelona Postgrad Encounters on Fundamental Physics, [arXiv:1302.0532 [gr-qc]]

\bibitem{NonLinearHorndeski} S. A. Appleby, A. De Felice, and E. V. Linder, JCAP {\bf 1210}, 060 (2012), [arXiv:1208.4163 [astro-ph.CO]]


\bibitem{JCAP-Alvaro} 
  F.~D.~Albareti, J.~A.~R.~Cembranos, A.~de la Cruz-Dombriz and A.~Dobado,   %``On the non-attractive character of gravity in f(R) theories,''
  JCAP {\bf 1307} (2013) 009   [arXiv:1212.4781 [gr-qc]];
  B.~Jain, V.~Vikram and J.~Sakstein, %``Astrophysical Tests of Modified Gravity: Constraints from Distance Indicators in the    Nearby Universe,''
  Astrophys.\ J.\  {\bf 779} (2013) 39   [arXiv:1204.6044 [astro-ph.CO]];
  L.~Lombriser, A.~Slosar, U.~Seljak and W.~Hu,   %``Constraints on f(R) gravity from probing the large-scale structure,''
  Phys.\ Rev.\ D {\bf 85}, 124038 (2012) [arXiv:1003.3009 [astro-ph.CO]]; 
  L.~Lombriser, F.~Schmidt, T.~Baldauf, R.~Mandelbaum, U.~Seljak and R.~E.~Smith,   %``Cluster Density Profiles as a Test of Modified Gravity,''
  Phys.\ Rev.\ D {\bf 85}, 102001 (2012) [arXiv:1111.2020 [astro-ph.CO]];
  Y.~-S.~Song, H.~Peiris and W.~Hu,   %``Cosmological Constraints on f(R) Acceleration Models,''
  Phys.\ Rev.\ D {\bf 76}, 063517 (2007) [arXiv:0706.2399 [astro-ph]];
  Y.~-S.~Song, W.~Hu and I.~Sawicki,   %``The Large Scale Structure of f(R) Gravity,''
  Phys.\ Rev.\ D {\bf 75}, 044004 (2007) [astro-ph/0610532].


%\cite{Poplawski:2006ey}
\bibitem{Poplawski:2006ey}
  N.~J.~Poplawski,
  %``A Lagrangian description of interacting dark energy,''
  gr-qc/0608031.
  %%CITATION = GR-QC/0608031;%%
  %27 citations counted in INSPIRE as of 23 Feb 2015
  \bibitem{fRTpaper} T. Harko, F. S. N. Lobo, S. Nojiri and S. D. Odintsov, Phys. Rev. {\bf D84} (2011) 024020. [arXiv:1104.2669 [gr-qc]].

 
\bibitem{Varia_fRT} %\cite{Sharif:2014ioa}
%\bibitem{Sharif:2014ioa}
  M.~Sharif and M.~Zubair,  %``Cosmological reconstruction and stability in $f(R,T)$ gravity,''
  Gen.\ Rel.\ Grav.\  {\bf 46} (2014) 1723; 
  %\cite{Sharif:2014jpa}
%\bibitem{Sharif:2014jpa}
  % M.~Sharif and M.~Zubair %``Reconstruction and stability of $\mathcal{f}(R,T)$ gravity with Ricci and modified Ricci dark energy,''
  Astrophys.\ Space Sci.\  {\bf 349} (2014) 529.
%\cite{Shabani:2013djy} \bibitem{Shabani:2013djy}
  H.~Shabani and M.~Farhoudi, %``f(R,T) Cosmological Models in Phase Space,''
  Phys.\ Rev.\ D {\bf 88} (2013) 044048   [arXiv:1306.3164 [gr-qc]];
%  
 T.~Harko and F.~S.~N.~Lobo, %``Generalized curvature-matter couplings in modified gravity,''
  Galaxies 2 (2014) 3, 410-465   [arXiv:1407.2013 [gr-qc]];
  %\cite{Harko:2014pqa}
  T.~Harko,   %``Thermodynamic interpretation of the generalized gravity models with geometry - matter coupling,''
  arXiv:1408.3465 [gr-qc].
  
  
\bibitem{stephaneseul3} M. J. S. Houndjo, Int. J. Mod. Phys. D. {\bf 21}, 1250003 (2012). arXiv: 1107.3887 [astro-ph.CO];
%\bibitem{oliver} 
M. J. S. Houndjo and O. F. Piattella,  Int. J. Mod. Phys. D. {\bf 21},   1250024  (2012). arXiv: 1111.4275 [gr.qc];
%\bibitem{momeni} 
D. Momeni, M. Jamil and R. Myrzakulov, Euro. Phys. J. C {\bf 72}, arXiv: 1107.5807 [physics.gen-ph].

\bibitem{flavio} F. G. Alvarenga, M. J. S. Houndjo, A. V. Monwanou and Jean. B. Chabi-Orou, %``Testing some f(R, T) gravity models from energy conditions" 
Journal of Modern Physics {\bf 4}, 130-139 (2013). %arXiv: 1205.4678 [gr-qc].

\bibitem{thermo1} M. Sharif and M. Zubair, JCAP {\bf 03}, 028 (2012); arXiv:1204.0848v2 [gr-qc].
%\bibitem{Jamil:2012pf}
  M.~Jamil, D.~Momeni and R.~Myrzakulov,
  %``Violation of First Law of Thermodynamics in f(R,T) Gravity,''
  Chin.\ Phys.\ Lett.\  {\bf 29}, 109801 (2012)
  [arXiv:1209.2916 [physics.gen-ph]].
  %%CITATION = ARXIV:1209.2916;%%

\bibitem{Harko_August_2014} T. Harko, %Thermodynamic interpretation of the generalized gravity models with geometry - matter coupling
	arXiv:1408.3465 [gr-qc]
	
\bibitem{juliano} M. J. S. Houndjo, C. E. M. Batista, J. P. Campos and O. F. Piattella, % `Finite-timne singularities in $f(R, T)$ gravity and the effect of conformal anomaly''. 
Can. J. Phys. {\bf 91} (7), 548-553 (2013). % arXiv:1203.6084 [gr-qc].

\bibitem{Nesseris:2008mq} 
  S.~Nesseris,
  %``Matter density perturbations in modified gravity models with arbitrary coupling between matter and geometry,''
  Phys.\ Rev.\ D {\bf 79}, 044015 (2009).

\bibitem{Diego-Alvaro-PRD2013} %\cite{Alvarenga:2013syu} \bibitem{Alvarenga:2013syu}
  F.~G.~Alvarenga, A.~de la Cruz-Dombriz, M.~J.~S.~Houndjo, M.~E.~Rodrigues and D.~S\'aez-G\'omez,  %``Dynamics of scalar perturbations in f(R,T) gravity,''
  Phys.\ Rev.\ D {\bf 87} (2013) 10,  103526    [arXiv:1302.1866 [gr-qc]].
  
  
  
\bibitem{Odintsov_HL} 
S. Nojiri and S. D. Odintsov, Phys. Rev. D {\bf 81} (2010) 043001 [arXiv:0905.4213 [hep-th]];Phys. Rev. D {\bf 83} (2011) 023001 [arXiv:1007.4856 [hep-th]].


  %\cite{Sharif:2013kga}
\bibitem{Sharif:2013kga}
  M.~Sharif and M.~Zubair,   %``Energy Conditions in $f(R,T,R_{\mu\nu}T^{\mu\nu})$ Gravity,''
  JHEP {\bf 1312} (2013) 079   [arXiv:1306.3450 [gr-qc]].



\bibitem{Thermo-Zubair} % Study of thermodynamic laws in f(R,T,R??T??) gravity
M. Sharif, M. Zubair, JCAP {\bf 1311} (2013) 042.

%\cite{Barcelo:2014mua}
\bibitem{Barcelo:2014mua}
  C.~Barcel\'o, R.~Carballo-Rubio and L.~J.~Garay,
  %``Unimodular gravity and general relativity from graviton self-interactions,''
  Phys.\ Rev.\ D {\bf 89} (2014) 124019
  [arXiv:1401.2941 [gr-qc]].
  %%CITATION = ARXIV:1401.2941;%%
  %5 citations counted in INSPIRE as of 16 Oct 2014


%\cite{Singh:1988pp}
% \bibitem{Singh:1988pp}
 % T.~P.~Singh and T.~Padmanabhan,   %``An Attempt to Explain the Smallness of the Cosmological Constant,''
 % Int.\ J.\ Mod.\ Phys.\ A {\bf 3} (1988) 1593.
%\cite{Sami:2002se}
%\bibitem{Sami:2002se}
 % M.~Sami and T.~Padmanabhan,   %``A Viable cosmology with a scalar field coupled to the trace of the stress tensor,''
 % Phys.\ Rev.\ D {\bf 67} (2003) 083509  [Erratum-ibid.\ D {\bf 67} (2003) 109901]  [hep-th/0212317].

\bibitem{fR varia}   
  A.~de~Felice and S.~Tsujikawa, Living Rev. Relativity {\bf 13}, 3 (2010). 
  T.~P.~Sotiriou, J.\ Phys.\ Conf.\ Ser.\  {\bf 189}, 012039 (2009); %[arXiv:0810.5594 [gr-qc]].
  S.~Capozziello and M.~De Laurentis, Phys.\ Rept.\  {\bf 509}, 167 (2011); %[arXiv:1108.6266 [gr-qc]].
  S.~'i.~Nojiri and S.~D.~Odintsov, Phys.\ Rept.\  {\bf 505}, 59 (2011); %[arXiv:1011.0544 [gr-qc]].
  T.~P.~Sotiriou and V.~Faraoni, Rev.\ Mod.\ Phys.\  {\bf 82}, 451 (2010);  %[arXiv:0805.1726 [gr-qc]].
  A.~de la Cruz-Dombriz and D.~S\'aez-G\'omez, Entropy {\bf 14}, 1717 (2012);
  A.~de la Cruz-Dombriz and A.~Dobado,   %``A f(R) gravity without cosmological constant,''
  Phys.\ Rev.\ D {\bf 74}, 087501 (2006)   [gr-qc/0607118];
  %\cite{Dunsby:2010wg}
  P.~K.~S.~Dunsby, E.~Elizalde, R.~Goswami, S.~Odintsov and D.~Saez-Gomez,  %``On the LCDM Universe in f(R) gravity,''
  Phys.\ Rev.\ D {\bf 82}, 023519 (2010)  [arXiv:1005.2205 [gr-qc]]; 
%\cite{Nojiri:2006be}
  S.~'i.~Nojiri and S.~D.~Odintsov, %``Modified gravity and its reconstruction from the universe expansion history,''
  J.\ Phys.\ Conf.\ Ser.\  {\bf 66}, 012005 (2007)   [hep-th/0611071];
  %\cite{Nojiri:2009kx}
  S.~'i.~Nojiri, S.~D.~Odintsov and D.~Saez-Gomez, %``Cosmological reconstruction of realistic modified F(R) gravities,''
  Phys.\ Lett.\ B {\bf 681}, 74 (2009)  [arXiv:0908.1269 [hep-th]];
  %\cite{Carloni:2010ph}
  S.~Carloni, R.~Goswami and P.~K.~S.~Dunsby,   %``A new approach to reconstruction methods in $f(R)$ gravity,''
  Class.\ Quant.\ Grav.\  {\bf 29}, 135012 (2012)  [arXiv:1005.1840 [gr-qc]];
%\cite{Goheer:2009ss}
  N.~Goheer, J.~Larena and P.~K.~S.~Dunsby,  %``Power-law cosmic expansion in f(R) gravity models,''
  Phys.\ Rev.\ D {\bf 80}, 061301 (2009)   [arXiv:0906.3860 [gr-qc]]
  

%\cite{Germani:2010hd}
\bibitem{Germani:2010hd}
  C.~Germani and A.~Kehagias,  %``UV-Protected Inflation,''
  Phys.\ Rev.\ Lett.\  {\bf 106} (2011) 161302   [arXiv:1012.0853 [hep-ph]].
 
  
%\cite{Rinaldi:2012vy}
\bibitem{Rinaldi:2012vy}
  M.~Rinaldi,   %``Black holes with non-minimal derivative coupling,''
  Phys.\ Rev.\ D {\bf 86} (2012) 084048 [arXiv:1208.0103 [gr-qc]].

%\cite{deRham:2011by}
\bibitem{deRham:2011by}
  C.~de Rham and L.~Heisenberg,
  %``Cosmology of the Galileon from Massive Gravity,''
  Phys.\ Rev.\ D {\bf 84} (2011) 043503
  [arXiv:1106.3312 [hep-th]];
  %%CITATION = ARXIV:1106.3312;%%
  %66 citations counted in INSPIRE as of 06 Nov 2014
%\cite{Heisenberg:2014kea}
%\bibitem{Heisenberg:2014kea}
  L.~Heisenberg, R.~Kimura and K.~Yamamoto,
  %``Cosmology of the proxy theory to massive gravity,''
  Phys.\ Rev.\ D {\bf 89} (2014) 103008
  [arXiv:1403.2049 [hep-th]].
  %%CITATION = ARXIV:1403.2049;%%
  %9 citations counted in INSPIRE as of 06 Nov 2014
%-------------------

%\bibitem{Koivisto2006}



  
%%%%%%%% f(R,T,RmmTmm)

% DK instability




%\cite{Horndeski:1976gi}
\bibitem{Horndeski:1976gi}
  G.~W.~Horndeski,
  %``Conservation of Charge and the Einstein-Maxwell Field Equations,''
  J.\ Math.\ Phys.\  {\bf 17} (1976) 1980.
  %%CITATION = JMAPA,17,1980;%%
  %31 citations counted in INSPIRE as of 30 Oct 2014

%\cite{Barrow:2012ay}
\bibitem{Barrow:2012ay}
  J.~D.~Barrow, M.~Thorsrud and K.~Yamamoto,
  %``Cosmologies in Horndeski's second-order vector-tensor theory,''
  JHEP {\bf 1302} (2013) 146
  [arXiv:1211.5403 [gr-qc]].
  %%CITATION = ARXIV:1211.5403;%%
  %6 citations counted in INSPIRE as of 30 Oct 2014

  %\cite{Jimenez:2013qsa}
\bibitem{Jimenez:2013qsa}  
J.~Beltr\'an Jim\'enez, R.~Durrer, L.~Heisenberg and M.~Thorsrud,  %``Stability of Horndeski vector-tensor interactions,''
  JCAP {\bf 1310} (2013) 064 [arXiv:1308.1867 [hep-th]].

  %\cite{Heisenberg:2014rta}
\bibitem{Heisenberg:2014rta}
  L.~Heisenberg,
  %``Generalization of the Proca Action,''
  JCAP {\bf 1405} (2014) 015
  [arXiv:1402.7026 [hep-th]].
  %%CITATION = ARXIV:1402.7026;%%
  %12 citations counted in INSPIRE as of 30 Oct 2014
  
  %\cite{BeltranJimenez:2010uh}
\bibitem{BeltranJimenez:2010uh}
  J.~Beltr\'an Jimenez and A.~L.~Maroto,
  %``Dark energy, non-minimal couplings and the origin of cosmic magnetic fields,''
  JCAP {\bf 1012} (2010) 025
  [arXiv:1010.4513 [astro-ph.CO]].
  %%CITATION = ARXIV:1010.4513;%%
  %10 citations counted in INSPIRE as of 27 Aug 2014
  
  
  %\cite{Jimenez:2014rna}
\bibitem{Jimenez:2014rna}
  J.~Beltr\'an Jim\'enez and T.~S.~Koivisto,
  %``Extended Gauss-Bonnet gravities in Weyl geometry,''
  Class.\ Quant.\ Grav.\  {\bf 31} (2014) 135002
  [arXiv:1402.1846 [gr-qc]].
  %%CITATION = ARXIV:1402.1846;%%
  %2 citations counted in INSPIRE as of 29 Aug 2014


% \bibitem{Beyond_Horndeski} Beyond Horndeski

%\bibitem{Reconstruction}


\end{thebibliography}
\end{document}